\documentclass[twocolumn,epjc3]{svjour3}
\journalname{Eur. Phys. J. A}
\smartqed  

\usepackage[utf8]{inputenc}
\usepackage[T1]{fontenc}
\usepackage[english]{babel}
\usepackage[numbers,sort&compress]{natbib}

\usepackage{microtype}
\usepackage[parfill]{parskip}

\usepackage{amsmath}
\usepackage{amssymb}
\usepackage{mathabx}
\usepackage{bbm,url}
\usepackage{graphicx}
\usepackage{grffile}

\usepackage[usenames,dvipsnames,svgnames,table]{xcolor}

\usepackage{float}
\usepackage{caption}
\usepackage{subcaption}
\usepackage{verbatim}
\usepackage{afterpage}

\usepackage{simplewick}
\newcommand{\wca}[2]{\ensuremath{\contraction[1.5ex]{}{\aO}{{}^\dag_{#1}}{\aO}\aaO_{#1}\aO_{#2}}}
\newcommand{\wac}[2]{\ensuremath{\contraction[1.5ex]{}{\aO}{{}_{#1}}{\aO}\aO_{#1}\aaO_{#2}}}

\DeclareMathSymbol{\NS}{\mathord}{AMSb}{"4E}

\renewcommand{\Re}{\operatorname{Re}}

\newcommand{\ket}[1]{\ensuremath{\,|{#1}\rangle}}
\newcommand{\bra}[1]{\ensuremath{\langle{#1}}|\,}
\newcommand{\braket}[2]{\ensuremath{\langle{#1}|{#2}\rangle}}
\newcommand{\matrixe}[3]{\ensuremath{\langle{#1}|\,{#2}\,|{#3}\rangle}}

\newcommand{\expect}[1]{\ensuremath{\langle{#1}\rangle}}

\newcommand{\acomm}[2]{\ensuremath{ \big\{ {#1}, {#2} \big\} }}

\renewcommand{\vec}[1]{\ensuremath{\mathbf{#1}}}

\newcommand{\op}[1]{\ensuremath{#1}}
\newcommand{\adj}[1]{\ensuremath{{{#1}}^{\dag}}}

\newcommand{\nord}[1]{\ensuremath{\mathopen:{#1}\mathclose:}}

\newcommand{\aO}{\ensuremath{\op{a}}}

\newcommand{\aaO}{\ensuremath{\adj{\op{a}}}}

\newcommand{\eMax}{\ensuremath{e_\text{max}}}
\newcommand{\lMax}{\ensuremath{l_\text{max}}}

\newcommand{\EMax}{\ensuremath{E_\text{3,max}}}
\newcommand{\nuc}[2]{\ensuremath{{}^{#2}\mathrm{#1}}}
\newcommand{\MeV}{\ensuremath{\mathrm{MeV}}}
\newcommand{\fm}{\ensuremath{\mathrm{fm}}}

\newcommand{\ctot}{\ensuremath{c_\mathrm{tot}}}
\newcommand{\Eref}{\ensuremath{E_\text{ref}}}
\newcommand{\Href}{\ensuremath{H_\text{ref}}}
\newcommand{\Hpert}{\ensuremath{H_\text{pert}}}
\newcommand{\Ntot}{\ensuremath{N_\mathrm{tot}}}

\renewcommand{\epsilon}{\varepsilon}

\newtheorem{lem}{Lemma}

\newcommand{\tcomp}{v}
\newcommand{\Tcomp}{V}

\graphicspath{ {./Images/} }

\usepackage{hyperref}

\hypersetup{
    colorlinks = true,
    citecolor  = blue,
    linkcolor  = blue,
    urlcolor  = blue
}
\renewcommand{\email}[1]{e-mail: \href{mailto:#1}{#1}}

\hypersetup{draft}
\begin{document}

	\title{Modewise Johnson-Lindenstrauss Embeddings for Nuclear Many-Body Theory} 

    \author{A. Zare\thanksref{e1,addr1}
            \and
            R. Wirth\thanksref{e2,addr2} 
            \and
            C. A. Haselby\thanksref{e3,addr3} 
            \and
            H. Hergert\thanksref{e4,addr2,addr4} 
            \and
            M. Iwen\thanksref{e5,addr3,addr1} 
    }

    \thankstext{e1}{\email{zareali@msu.edu}}
    \thankstext{e2}{\email{wirth@frib.msu.edu}}
    \thankstext{e3}{\email{haselbyc@msu.edu}}
    \thankstext{e4}{\email{hergert@frib.msu.edu}}
    \thankstext{e5}{\email{iwenmark@msu.edu}}
    
    
    \institute{Department of Computational Mathematics, Science and Engineering, Michigan State University, East Lansing, MI 48824, USA \label{addr1}
               \and
               Facility for Rare Isotope Beams, Michigan State University, MI 48824, USA \label{addr2}
               \and Department of Mathematics, Michigan State University, East Lansing, MI 48824, USA \label{addr3}
               \and
              Department of Physics \& Astronomy, Michigan State University, MI 48824, USA \label{addr4}
    }
    
    \date{Received: \today{} / Accepted: date}

	\maketitle
	\begin{abstract}
		In the present work, we initiate a program that explores modewise Johnson-Lindenstrauss embeddings (JLEs) as a tool to reduce the computational cost and memory requirements of (nuclear) many-body methods. These embeddings are randomized projections of high-dimensional data tensors onto low-dimensional subspaces that preserve structural features like norms and inner products.

    An appealing feature of randomized embedding techniques is that they allow for the oblivious and incremental compression of large tensors, e.g., the nuclear Hamiltonian or wave functions amplitudes, into significantly smaller random sketches that still allow for the accurate calculation of ground-state energies and other observables. In particular, the oblivious nature of randomized JLE techniques makes it possible to compress a tensor without knowing in advance exactly what observables one might want to approximate at a later time. This opens the door for the use of tensors that are much too large to store in memory, e.g., untruncated three-nucleon forces in current approaches, or complete two- plus three-nucleon Hamiltonians in large, symmetry-unrestricted bases. Such compressed Hamiltonians can be stored and used later on with relative ease.

    As a first step, we perform a detailed analysis of a JLE's impact on the second-order Many-Body Perturbation Theory (MBPT) corrections for nuclear ground-state observables like the energy and the radius, noting that these will be the dominant corrections in a well-behaved perturbative expansion, and highly important implicit contributions even in nonperturbative approaches. Numerical experiments for a wide range of closed-shell nuclei, model spaces and state-of-the-art nuclear interactions demonstrate the validity and potential of the proposed approach: We can compress nuclear Hamiltonians hundred- to thousandfold while only incurring mean relative errors of 1\% or less in ground-state observables. 

    Importantly, we show that JLEs capture the relevant physical information contained in the highly structured Hamiltonian tensor despite their random characteristics. In addition to the significant storage savings, the achieved compressions imply multiple order-of magnitude reductions in computational effort when the compressed Hamiltonians are used in higher-order MBPT or nonperturbative many-body methods.
				
	\end{abstract}
	
\section{\label{sec:intro}Introduction}
The quantum many-body problem is a prime example of a data-intensive problem with relevance 
in the fundamental and applied sciences. The structure and dynamics of quantum many-body systems 
are governed by the stationary and time-dependent Schr\"odinger equations, respectively. In numerical 
simulations, they can be cast in the form of matrix eigenvalue or differential equations in straightforward 
fashion, but the dimension of the involved matrices grows exponentially with the number of particles 
and their degrees of freedom. 

Nowadays, exact diagonalization methods for atomic nuclei like the (No-Core) Shell Model (NCSM) or 
(No-Core) Full Configuration Interaction (NCFC) \cite{Barrett:2013oq,Navratil:2016rw} can tackle 
dimensions on the order of 10 billion because the natural scales and symmetries of the interactions 
induce a high degree of 
sparsity in the involved matrices \cite{Yang:2013ly,Hergert:2020am}. Still, the memory requirements
and computational cost can only be met by supercomputers, and the numerically tractable dimensions 
are merely sufficient to obtain results for nuclei up to mass $\sim20$ in this way, excluding thousands 
of isotopes that are predicted to exist in nature --- many of which will be produced for the first 
time under laboratory conditions by rare-isotope facilities like the recently launched Facility for 
Rare Isotope Beams (FRIB) \cite{Balantekin:2014gf}.

To overcome the ``curse of dimensionality'' that plagues exact diagonalization approaches, one can 
deploy methods that solve the Schr\"odinger equation with systematic approximations \cite{Hergert:2020am}, 
e.g., Many-Body Perturbation Theory (MBPT) based on mean-field wave functions, or sophisticated 
nonperturbative approaches like the In-Medium Similarity Renormalization Group (IMSRG) 
\cite{Hergert:2016jk,Hergert:2017kx,Stroberg:2019th}, Coupled Cluster (CC) theory \cite{Shavitt:2009,Hagen:2014ve},
or Self-Consistent Green's Function Methods \cite{Soma:2020lo,Barbieri:2022lj,Rios:2020jz}. 
These methods typically scale polynomially with the dimension $N$ of the single-particle basis that 
defines the degrees of freedom for individual nucleons, and only indirectly with the particle number. 
For example, the IMSRG at the commonly used IMSRG(2) truncation level naively scales as $O(N^6)$, 
whereas the more precise next-level truncations IMSRG(3) requires $O(N^9)$ efffort.

While greatly extending the range of tractable nuclei \cite{Hergert:2020am,Stroberg:2021qu,Hu:2022sw}, 
approximate many-body methods become highly data intensive themselves as we strive for 
greater precision or add degrees of freedom to the single-particle basis. Both efforts are necessary 
as applications seek to describe exotic nuclei that exhibit deformation and weak-binding effects.
They can result in ten- to hundredfold increases in $N$, which in turn increase the memory requirements 
and computational cost of many-body calculations by several orders of magnitude, rendering them infeasible 
with current and next-generation computing resources. 

To some extent, the need for large basis dimensions is driven by competing requirements
of common many-body frameworks. The two- and three-nucleon interactions that govern nuclear
structure and dynamics have compact representations in terms of the relative coordinates,
momenta, and spins of the interacting particles, but we can only formulate many-body wave 
function bases in terms of these degrees of freedom in very light nuclei \cite{Navratil:2000hf,Nogga:2006yg,Nogga:2000tu}).
For mass numbers $A\gtrsim 5$, the computational effort for constructing such wave functions
becomes unfeasibly large. Instead, we adopt a formulation based on independent-particle
states, the so called Slater determinants. They are easy to construct because they are
simple antisymmetrized products of single-particle wave functions, but not attuned to
the description of the correlations that are induced by nuclear interactions. Consequently,
an exponentially large basis of Slater determinants is required to capture those correlations.
Since interactions still only involve two or three nucleons, this implies a high
degree of redundancy in the matrix representations of the interaction operators 
because the remaining ``spectator'' nucleons can be in exponentially many 
configurations. Approximate many-body methods like CC and IMSRG are efficient 
because they explicitly address part of this redundancy, but they also eventually run out of steam.

To tackle the redundancy problem,
one can attempt to identify the principal components of the interactions and wave
functions, and leverage the resulting factorizations to change the computational
scaling of the targeted many-body methods. In quantum chemistry, for example, efforts 
to construct factorized CC methods have come to fruition in recent years 
\cite{Schutski:2017xr,Parrish:2019zg,Hohenstein:2019jx,Hohenstein:2022oj,Lesiuk:2019wf,Lesiuk:2020hq,Lesiuk:2021gm},
and the adoption of a tensorial viewpoint that better reflects the product nature
of the many-body wave functions has proven particularly useful. Similar efforts
have been launched in nuclear physics \cite{Zhu:2021gd,Tichai:2021xw,Tichai:2019xz,Tichai:2022rq},
but they face their own particular challenges: While nuclei have simpler geometries
than molecules, the two- and three-nucleon interactions have a much more complex
structure than the Coulomb interaction governing atomic and molecular systems.

A less ambitious approach is to attempt a reduction of the single-particle basis
size through an optimization of the orbitals, so that the relevant physics can be
captured with fewer degrees of freedom. The design of such optimized basis sets 
has a long and successful history in quantum chemistry --- see for a recent
review \cite{Nagy:2017ew}. In nuclear physics, it is much less common: There is
a strong preference for using a basis (spherical) HO states at least initially
because it allows an exact separation of the center-of-mass and relative degrees 
of freedom in few- and many-body states, provided the basis is truncated appropriately 
(see, e.g., the discussion in \cite{Hergert:2016jk} and references therein). 

Recent works have demonstrated that the eigenstates of the one-body density matrix 
with perturbative corrections through second order greatly accelerate the convergence 
of NCSM/NCFC, IMSRG and CC calculations \cite{Tichai:2019to,Hoppe:2021ij,Novario:2020gr}. 
This implies that the relevant physical information contained in the nuclear Hamiltonian
can be compressed from the original working basis into a much smaller natural orbital 
basis with controllable accuracy.

In the present work, we explore an alternative compression approach that is 
based on the seminal work of Johnson and Lindenstrauss. Their famous lemma 
\cite{Johnson:1984ch} proves that random projections of high-dimensional data 
into lower dimensional subspaces will preserve structures of the data set like
distances and inner products  with a high likelihood. Since its publication, it 
has become an important ingredient
for algorithms and data analysis workflows because it mitigates the exponential
growth of data sets \cite{Halko:2011cl,Mahoney:2011wj,Woodruff:2014nk,doi:10.1137/17M1112303,sun2020sketching}. 

Modern data science and machine learning have embraced tensorial representations
of data for their efficiency, and many of the standard methods and algorithms for
data analysis have been extended to tensors as a result \cite{Kolda:2009te}.
This includes the development of so-called modewise JL embeddings (JLEs) \cite{Iwen:2021my}
that are the focus of this work. As mentioned above, tensor-based methods are 
a natural match for the product structure of the many-body Hilbert space and they
allow for computation and memory-efficient implementations of standard operations.
Modewise embeddings are very compact and easy to store, hence they can be readily
used to embed additional tensors, e.g., for other observables of interest, in a 
compatible format in future applications.

In the following, we will apply modewise JLEs to the evaluation of nuclear ground-state
observables like the binding energy and mean-square radius in Many-Body Perturbation Theory. 
Our main goal is to develop a detailed understanding of the JLE's impact
on second-order MBPT, or MBPT(2), which aims to capture leading-order
correlations beyond the mean field (i.e., independent particle) description of atomic
nuclei. In a well-behaved MBPT expansion, MBPT(2) will give the dominant corrections
to nuclear observables, so this analysis is an important foundation for future
applications of the JLE to higher-order MBPT.

Since we know a priori which entries of the Hamiltonian tensor contribute to MBPT(2),
the ground-state energy correction can in principle be implemented with a storage 
cost of $O(N_o^2N_u^2)$, where $N_o\ll N$ and $N_u = N - N_o$ are the number of 
occupied and unoccupied single-particle states, respectively. We will demonstrate that
the modewise JLEs allow us to construct compressed versions of the \emph{full} Hamiltonian,
so called random sketches, that are competitive in size with this ``optimal'' storage requirement, at the cost of 
a small but controllable loss of accuracy for the MBPT energy and wave function. Importantly, the
JLEs are oblivious in the sense that they do not rely on any prior assumption about
the structure of the Hamiltonian at all, yet they are still able to capture the most 
relevant physical information. Moreover, the compressed Hamiltonian can be readily
stored and used as input for future applications, e.g., in higher-order MBPT or 
nonperturbative methods. In such calculations, even modest compressions of the
Hamiltonian will also enable order-of-magnitude savings in computational effort.

This work is organized as follows: In Section \ref{sec:JLbasics}, we will briefly 
discuss the basic ingredients of modewise JLEs. Section \ref{sec:mbpt}
reviews the MBPT formalism, with special emphasis on
the use of normal-ordered operators and a brief discussion of expectation values
for general observables. In Section \ref{sec:JLapplication}, we reformulate
MBPT corrections in terms of inner products and derive the expressions for 
applying one- and two-stage JL embeddings. Numerical results from applications
in closed-shell nuclei are analyzed in depth in Section \ref{sec:JLresults}, and
we conclude with a summary and outlook at the next stage of JL applications in
Section \ref{sec:conclude}. For completeness, we compile numerical results
for additional nuclei in 
\ref{app:results}.

  \section{Johnson-Lindenstrauss Embeddings}
    \label{sec:JLbasics}
	\subsection{Low-Rank Tensors} In this section, a brief introduction to low-rank tensors as well as the dimension reduction of tensors using JLEs is presented.
	\subsubsection{CANDECOMP/PARAFAC Decomposition}
	The Canonical Polyadic Decomposition, also known as CANDECOMP or PARAFAC, and abbreviated here as CPD, decomposes a tensor $X$ into a (weighted) sum of rank-$1$ tensors \cite{Kolda:2009te}. For $X \in \mathbb{R}^{n_1 \times \ldots \times n_d}$,
	\begin{equation}
		X \approx \hat{X} = \sum\limits_{k=1}^{r}g_k~ \mathbf{\tcomp}^{(1)}_{k} \circ \mathbf{\tcomp}^{(2)}_{k} \circ \dots \circ \mathbf{\tcomp}^{(d)}_{k},
		\label{equ:CP_sum}
	\end{equation}
	where $\circ$ denotes the outer product. The vector $\mathbf{\tcomp}^{(j)}_{k} \in \mathbb{R}^{n_j}$ can be considered as the $k$\textsuperscript{th} column in a matrix $\mathbf{\Tcomp}^{(j)} \in \mathbb{R}^{n_j \times r}$ for $j \in [d]$.\footnote{We will use the notation $[d]:=\{1,\dots,d\}$ throughout the paper.} When solving for $\mathbf{\Tcomp}^{(j)}$ using a CPD fitting algorithm, the columns do not necessarily have unit norms, and so the columns can be normalized and the norms can be stored as weights $g_k=\prod_{j=1}^{d}\|\mathbf{\tcomp}_k^{(j)}\|_2$. However, if this normalization is not done, one can assume that all coefficients $g_k$ are one.
	
	\subsubsection{Tensor Rank} The rank of a tensor $X$ is defined as the smallest number of rank-$1$ tensors that generate $X$ as their sum. In other words, it is the smallest number of components in an {\it exact} CPD. In the context of Eq.~\eqref{equ:CP_sum}, the rank $r$ of a tensor $X$ is defined as the minimum rank $r$ such that $X = \hat{X}$ holds.
	Although the definition of tensor rank is analogous to matrix rank, the properties of the two ranks are very different from each other. A major difference is that computation of the rank of a tensor is known to be NP-complete \cite{Hastad1990}. Therefore, in practice, it is determined numerically by fitting various CP models.

\subsection{Johnson-Lindenstrauss Embeddings for Tensor Dimension Reduction}
Johnson-Lindenstrauss embeddings provide a simple yet powerful tool for dimension reduction of high-dimensional data using random linear projections. The following definition and lemma for matrices ($2$-mode tensors) is the fundamental building block used to extend results about random projections to tensors of any number of modes with low-rank structure.
\begin{definition}
    \label{def:jl_matrices}
	A matrix $\mathbf{A} \in \mathbb{C}^{m \times n}$ is an $\epsilon$-JL embedding of a set $S \subset \mathbbm{C}^n$ into $\mathbbm{C}^m$ if
	\begin{equation}
		\begin{split}
			\| \mathbf{A}\mathbf{x} \|_2^2=\left( 1+\epsilon_{\mathbf{x}} \right)\| \mathbf{x} \|_2^2,
		\end{split}
		\label{equ:JL_lem}
	\end{equation}
	\noindent with  $\left| \epsilon_{\mathbf{x}} \right| \varleq \epsilon$ for all $\mathbf{x} \in S$.
\end{definition}
Assuming that the elements of $\mathbf{A}$ are independent subgaussian random variables with mean zero and variance $m^{-1}$ and that $\left| S \right| =M$, then Eq. \eqref{equ:JL_lem} holds for all $\mathbf{x} \in S$ with probability $p \vargeq 1-2\exp{\left(-Cm\epsilon^2\right)}$ if $m \vargeq C\epsilon^{-2}\log{M}$, where $C$ is an absolute constant \cite{vershynin2018high}.

\begin{lem}
	Suppose that $X, Y \in \mathbb{R}^{n_1 \times \dots \times n_d}$ are rank-$r$ tensors of the form 
	\begin{equation}
	  X = \sum_{k=1}^r \alpha_k~ \mathbf{\tcomp}^{(1)}_{k} \circ \mathbf{\tcomp}^{(2)}_{k} \circ \dots \circ \mathbf{\tcomp}^{(d)}_{k},  
	\end{equation}
	and
	\begin{equation}
    	Y = \sum_{k=1}^r \beta_k~ \mathbf{\tcomp}^{(1)}_{k} \circ \mathbf{\tcomp}^{(2)}_{k} \circ \dots \circ \mathbf{\tcomp}^{(d)}_{k}.	    
	\end{equation}
	Let $\epsilon \in (0, 3/4]$, and $\mathbf{A}^{(j)} \in \mathbb{R}^{m_j \times n_j}$ be a $(\epsilon / (4d) )$-JL embedding for each $j \in [d]$.  
	Then,
	\begin{align}
		\left| \left\langle X \bigtimes_{j = 1}^d \mathbf{A}^{(j)},~ Y \bigtimes_{j = 1}^d \mathbf{A}^{(j)} \right\rangle - \left\langle X,~ Y\right\rangle \right|\notag\\ 
		~\leq C  \max \left\{ \| X \|^2, \| Y \|^2 \right\},
		\label{eq:innerprod_JL}
	\end{align}
	\noindent where $C$ depends on $\varepsilon$, $r$, $d$ and properties of the space spanned by $\mathbf{\tcomp}^{(1)}_{k} \circ \mathbf{\tcomp}^{(2)}_{k} \circ \dots \circ \mathbf{\tcomp}^{(d)}_{k}$. The complete version of this lemma can be found in Corollary $1$ in \cite{Iwen:2021my}.
		
	\label{lem:BasicVecInnProdJL}
\end{lem}

\noindent Here, we have defined the inner product of two tensors $X$ and $Y$ in terms of their components as\footnote{The set of all $d$-mode tensors $X \in \mathbbm{C}^{n_1 \times \ldots \times n_d}$ forms a vector space over the field of complex numbers when equipped with component-wise addition and scalar multiplication. Further equipping this vector space with an inner product operation leads to a standard inner product space of tensors.}
\begin{equation}
    \langle X,Y \rangle := \sum_{j_1\ldots j_d} X_{j_1\ldots j_d}\overline{Y_{j_1\ldots j_d}} \,.
    \label{eq:def_inner_prod}
\end{equation}

In Eq. \eqref{eq:innerprod_JL}, $X\times_j\mathbf{A}^{(j)}$ denotes the $j$-mode product between $X$ and $\mathbf{A}^{(j)}$, which is calculated by contracting the mode-$j$ fibers of $X$ with the rows of $\mathbf{A}^{(j)}$. Alternatively, one can write $\left( X\times_j\mathbf{A}^{(j)} \right)_{(j)}=\mathbf{A}^{(j)}\mathbf{X}_{(j)}$ where $\mathbf{X}_{(j)}$ is the mode-$j$ unfolding of $X$. Component-wise,
\begin{equation}
    \left( X\times_j\mathbf{A}^{(j)} \right)_{i_1 \ldots i_{j-1}\ell i_{j+1}\ldots i_d}=\sum\limits_{i_j=1}^{n_j} \mathbf{A}^{(j)}_{\ell i_j}X_{i_1 \ldots i_j \ldots i_d},
    \label{eq:j-mode_prod_single}
\end{equation}
for $\ell \in [m_j]$ and $i_j \in [n_j]$.
Extending the same idea to multiple modes, the components of the fully projected tensor can be calculated using
\begin{equation}
    \left(X \bigtimes_{j = 1}^d \mathbf{A}^{(j)} \right)_{\ell_1 \ldots  \ell_d}\!\!=\!\sum\limits_{i_1=1}^{n_1}\!\dots\!\! \sum\limits_{i_d=1}^{n_d} X_{i_1 \ldots i_d}\mathbf{A}^{(j)}_{\ell_1 i_1}\dots \mathbf{A}^{(d)}_{\ell_d i_d},
    \label{eq:j-mode_prod_full}
\end{equation}
for $\ell_j \in [m_j]$ and all $j \in [d]$. 
By applying JL embeddings to a low-rank $d$-mode tensor $X$, the size of all modes can be reduced to yield a projected tensor of much smaller size (a so-called random sketch), without the need to reshape the tensor into a single large vector \footnote{For information on mode-$j$ fibers and unfoldings, and mode products in tensors, see \cite{Kolda:2009te}.}. It is expected then that the Euclidean norm of the projected tensor as well as the inner product between any two tensors lying on the same low-rank subspace are preserved to within predictable errors.
Working modewise as per Lemma \ref{lem:BasicVecInnProdJL} requires the storage of several small JL matrices, whereas vectorizing the tensor and then using a single matrix to embed it requires significantly more memory since the JL matrix has more entries than the original tensor itself.


\section{Many-Body Perturbation Theory}
\label{sec:mbpt}
In the present work, we use the framework of Many-Body Perturbation Theory as a test-bed for the application of JLEs to nuclear interactions and observables. Specifically, we will focus on second-order M\o{}ller-Plesset MBPT corrections to the energies and general observables as one of the simplest post-Hartree Fock approaches to capture correlations --- deploying the JLEs in nonperturbative techniques like the IMSRG \cite{Hergert:2016jk,Hergert:2017kx} or the CC method \cite{Shavitt:2009,Hagen:2014ve} is more challenging, and will be considered in future work.

\subsection{Choice of Reference State and Normal Ordering}
The starting point for MBPT is the choice of the reference state $\ket{\Phi}$ on which the perturbative expansion will be built. As in any expansion method, this choice will affect both the form and the convergence of the expansion, as discussed in Refs.~\cite{Langhammer:2012uq,Tichai:2016vl,Tichai:2020ft}, for example. Here, we follow the \emph{canonical} approach and select a Hartree-Fock Slater determinant that has been variationally optimized for a nucleus of interest as our reference. We will denote this state by $\ket{\Phi}$ in the following. This allows us to construct a complete basis for our many-body Hilbert space from all physically allowed excitations of $\ket{\Phi}$.

The Hamiltonian and other observables will be expressed in terms of Fermionic creation ($\aaO_p$) and annihilation ($\aO_p$) operators that satisfy the fundamental anticommutation relations
\begin{equation}\label{eq:cac}
    \acomm{\aO_p}{\aaO_q} = \delta_{pq}\,,\quad \acomm{\aO_p}{\aO_q} = \acomm{\aaO_p}{\aaO_q} = 0\,,
\end{equation}
where $\acomm{x}{y} = xy+yx$ and the indices refer to the single-particle states of the Hartree-Fock basis. For organizing the excited states of our nucleus, it is convenient to \emph{normal order} relevant operators with respect to the reference state $\ket{\Phi}$. We define a normal-ordered one body operator as
\begin{equation}\label{eq:def_nord}
    \nord{\aaO_p\aO_q} \equiv \aaO_p\aO_q - \contraction[1.5ex]{}{\aO}{{}^\dag_p}{\aO}\aaO_p\aO_q\,,
\end{equation}
where $\contraction[1.5ex]{}{\aO}{{}^\dag_p}{\aO}\aaO_p\aO_q$ is a so-called (Wick) contraction between a creation and an annihilation operator. For Slater determinant references, we have 
\begin{equation}
    \contraction[1.5ex]{}{\aO}{{}^\dag_p}{\aO}\aaO_p\aO_q \equiv \rho_{qp} \equiv \matrixe{\Phi}{\aaO_p\aO_q}{\Phi}\,,
\end{equation}
i.e., the contraction is simply given by the one-body density matrix of the nucleus. We can recursively extend the normal ordering to higher-body operators:
\begin{align}
    \aaO_p\aaO_q\aO_s\aO_r &= \nord{\aaO_p\aaO_q\aO_s\aO_r} \notag\\
    &\quad+
    \wca{q}{s}
    \nord{\aaO_p\aO_r} - \wca{p}{s} \nord{\aaO_q\aO_r} \notag\\
    &\quad+ \wca{p}{r} \nord{\aaO_q\aO_s} - \wca{q}{r} \nord{\aaO_p\aO_s}  \notag\\
    &\quad+\left(\wca{p}{r}\,\wca{q}{s} -\wca{p}{s}\,\wca{q}{r}\right)\,,\\
    \aaO_p\aaO_q\aaO_r\aO_u\aO_t\aO_s  &= \nord{\aaO_p\aaO_q\aaO_r\aO_u\aO_t\aO_s} \notag\\
    &\quad + \wca{r}{u}\nord{\aaO_p\aaO_q\aO_t\aO_s} + \ldots \notag\\
    &\quad + \left(\wca{r}{u}\,\wca{q}{t} - \wca{q}{u}\,\wca{r}{t} \right)\nord{\aaO_p\aO_s} + \ldots \notag\\
    &\quad + \wca{p}{s}\,\wca{q}{t}\,\wca{r}{u} + \ldots 
\end{align}

The definition \eqref{eq:def_nord} immediately implies that the expectation value of a normal-ordered one-body operator in the reference state vanishes, and it is straightforward to confirm that this holds for any normal-ordered operator:
\begin{equation}\label{eq:expect_nord}
    \matrixe{\Phi}{\nord{\aaO_p\ldots \aO_q}}{\Phi} = 0\,.
\end{equation}
This leads to significant simplifications when we evaluate expressions in MBPT or other many-body methods. In addition, any product of two normal-ordered operators can be expanded by contracting the creators and annihilators of one string with a suitable match of the other string, e.g.,
\begin{align}
    \nord{\aaO_{p}\aO_q}\nord{\aaO_r\aO_s} = \wac{q}{r}\nord{\aaO_p\aO_s} - \wca{p}{s}\nord{\aaO_r\aO_q} + \wca{p}{s}\wac{q}{r}\,,
\end{align}
where we have used that\footnote{The contraction explicitly accounts for the contribution of the nonvanishing RHS of the fundamental anticommutator Eq.~\eqref{eq:cac} so that we only need to track signs when we permute operators within a normal-ordered string.} $\nord{\aO_q\aaO_r} = -\nord{\aaO_r\aO_q}$, and introduced
\begin{equation}
    \wac{q}{r} = \delta_{qr} - \rho_{qr}\,.
\end{equation}
These are the essential rules for evaluating the MBPT expressions we are considering in the following. Additional details can be found in Refs.~\cite{Shavitt:2009,Hergert:2016jk,Kutzelnigg:1997fk,Hergert:2017kx}, for example.

\subsection{The Normal-Ordered Hamiltonian}
\label{sec:no}
Since nuclei are self-bound objects, the nuclear many-body Hamiltonian is constructed using the \emph{intrinsic kinetic energy}\footnote{In particle-number conserving approaches, it can equivalently be written as a sum of two-body operators alone, but this has a subtle impact on HF and MBPT --- see Refs.~\cite{Khadkikar:1974uv,Hergert:2009wh}.}
\begin{equation}
    T_\text{int} \equiv T - T_\text{cm} \equiv T_1 + T_2\,.
\end{equation}

Considering two- and three-nucleon interactions, which is standard nowadays, we have a starting Hamiltonian of the form
\begin{equation}
    H = T_1 + T_2 + V_2 +V_3\,.
\end{equation}
It can be written in normal ordered form as
\begin{align}
    H &= \Eref + \sum_{pq}H_{pq}\nord{\aaO_p\aO_q} + \frac{1}{4}\sum_{pqrs}H_{pqrs}\nord{\aaO_p\aaO_q\aO_s\aO_r}\notag\\
    &\quad+\frac{1}{36}\sum_{pqrstu}H_{pqrstu}\nord{\aaO_p\aaO_q\aaO_r\aO_u\aO_t\aO_s}\,,\label{eq:def_nord_H}
\end{align}
with
\begin{align}
    \Eref &= \matrixe{\Phi}{H}{\Phi}\,,\\
    H_{pq} &= T_{pq} + \sum_{rs}(T + V)_{prqs}\rho_{sr} 
     \notag\\
     &\quad + \frac{1}{2}\sum_{rstu}V_{prsqtu}\rho_{tr}\rho_{us}\,,\\
    H_{pqrs} &=(T+V)_{pqrs} + \sum_{tu} V_{pqtrsu}\rho_{ut}\,,\\
    H_{pqrstu} &= V_{pqrstu}\,.
\end{align}
Here, we assume that all two- and three-body matrix elements are fully antisymmetrized under permutations of the indices, e.g.,
\begin{align}
    O_{pqrs} = -O_{qprs}
    = -O_{pqsr} = O_{qpsr}\,\label{eq:perm}\,,
\end{align}
where $O = H,T,V$. Analogous relations hold for general observables.

Due to the normal ordering, the dominant effects of the three-nucleon interaction are absorbed into the zero-, one-, and two-body parts of the normal ordered Hamiltonian, and we can neglect the residual three-body terms in the following. This so-called normal-ordered two-body approximation (NO2B) that is frequently used in state-of-the-art nuclear many-body calculations. It significantly reduces the storage and computational effort at the cost of a typical approximation error of 1-2\% in medium-mass nuclei \cite{Hagen:2007zc,Roth:2012qf}.  

The normal-ordered Hamiltonian can be simplified if we use the HF orbitals of our reference state $\ket{\Phi}$ as our single-particle basis: Then
the one-body part of the Hamiltonian and the one-body density matrix are both diagonal,
\begin{align}
    H_{pq} &= \epsilon_p\delta_{pq}\,,\\
    \rho_{pq} &= n_p\delta_{pq}\,, 
\end{align}
and their eigenvalues define the single-particle energies $\epsilon_p$ and the occupation numbers $n_p$, respectively. In a HF state for a given nucleus with $Z$ protons and $N$ neutrons, the $Z$ proton and $N$ neutron states with the lowest single-particle energies are occupied ($n_p = 1$) while all other single-particle states are unoccupied ($n_p = 0$). In a typical calculation, the number of occupied states is much smaller than the number of unoccupied states, $N_o \ll N_u \sim N$, where $N = N_u + N_o$ is the single-particle  basis size. 

Adopting a convention from quantum chemistry, we will use indices $i,j,k,\ldots$ to label occupied (``hole'') states in subsequent expressions, while unoccupied (``particle'') states are indicated by $a,b,\ldots$, and $p,q,\ldots$ do not distinguish between the two types of states and encompass the entire single-particle basis.

\subsection{Energy Corrections}
Using the reference state as a basis, we can construct perturbative corrections to the wave function and energies in the usual M\o{}ller-Plesset scheme (see, e.g., \cite{Shavitt:2009,Tichai:2020hi}). We partition the Hamiltonian into 
\begin{equation}
    H = \Href + g \Hpert\,,
\end{equation}
where 
\begin{align}
    \Href &= \Eref + \sum_{p}\epsilon_p \nord{\aaO_p\aO_p}\,,\\
    \Hpert &= \frac{1}{4}\sum_{pqrs}H_{pqrs}\nord{\aaO_p\aaO_q\aO_s\aO_r}
\end{align}
and $g$ is merely a book-keeping parameter that will be set to 1 in practical calculations. Formally, the energy and the many-body state are then expanded as
\begin{align}
    E &= E^{(0)} + g E^{(1)} +g^2 E^{(2)} + \ldots \\
    \ket{\Psi} &= \ket{\Psi^{(0)}} + g\ket{\Psi^{(1)}} + g^2\ket{\Psi^{(2)}} + \ldots 
\end{align}
where the corrections $\ket{\Psi^{(k)}}$ for $k>0$ are supposed to be orthogonal to the leading-order wave function:
\begin{equation}\label{eq:intermediate_norm}
    \braket{\Psi^{(0)}}{\Psi^{(k)}} = 0\,.
\end{equation}
Plugging the expansions into the Schr\"odinger equation,
\begin{align}
    &(\Href + g\Hpert) \left(\ket{\Psi^{(0)}} + g\ket{\Psi^{(1)}} + \ldots\right) \notag\\
    &= (E^{(0)} + g E^{(1)} + \ldots)\left(\ket{\Psi^{(0)}} + g\ket{\Psi^{(1)}} + \ldots\right)\,,
\end{align}
and comparing powers of $g$, one immediately obtains
\begin{equation}
    \Href\ket{\Psi^{(0)}} = E^{(0)}\ket{\Psi^{(0)}} = E_\text{ref}\ket{\Psi^{(0)}}\,.
\end{equation}
The energy corrections can now be obtained by projecting onto the unperturbed state,
\begin{equation}
    E^{(k)} = \matrixe{\Psi^{(0)}}{\Hpert}{\Psi^{(k-1)}}\,,
\end{equation}
and the wave function corrections are given by
\begin{align}
    \ket{\Psi^{(k)}} &= \frac{Q}{\Eref-\Href} \Hpert \ket{\Psi^{(k-1)}} \notag\\
    &\qquad - \sum_{j=1}^{k-1} E^{(j)}\frac{ Q}{\Eref-\Href}\ket{\Psi^{(k-j)}}\,,
\end{align}
where the projection operator
\begin{equation}
    Q = 1 - \ket{\Psi^{(0)}}\bra{\Psi^{(0)}}
\end{equation}
was introduced to ensure that Eq.~\eqref{eq:intermediate_norm} holds at all orders.

In the canonical MBPT with a Hartree-Fock Slater determinant reference state $\ket{\Phi}$, we have
\begin{equation}
    E^{(0)} = E_\text{ref},\quad \ket{\Psi^{(0)}} = \ket{\Phi}\,.
\end{equation}
Because $\Hpert$ is in normal order, $E^{(1)} = \matrixe{\Phi}{\Hpert}{\Phi} = 0$ and the first correction to the energy appears at second order.
The corrections are expanded in terms of particle-hole excitations of the reference state:
\begin{equation}
    \ket{\Phi^{ab\ldots}_{ij\ldots}} = \aaO_a\aaO_b\ldots\aO_j\aO_i\ket{\Phi}
    = \nord{\aaO_a\aaO_b\ldots\aO_j\aO_i}\ket{\Phi}\,,
\end{equation}
where we have used that contractions between particle ($a,b,\ldots$) and hole indices ($i,j,\ldots)$ vanish. The vanishing expectation value of normal ordered operators in the reference state (see Eq.~\eqref{eq:expect_nord}) guarantees that the orthogonality condition \eqref{eq:intermediate_norm} is satisfied. Furthermore, the particle-hole excitations are eigenstates of $\Href$ with
\begin{align}
    H_\text{ref}\ket{\Phi^{ab\ldots}_{ij\ldots}} &= (E_\text{ref} + \epsilon_a + \epsilon_b + \ldots - \epsilon_i -\epsilon_j)\ket{\Phi^{ab\ldots}_{ij\ldots}}\notag\\
    &\equiv (E_\text{ref} + \epsilon^{ab\ldots}_{ij\ldots})\ket{\Phi^{ab\ldots}_{ij\ldots}}\,,\label{eq:eig_Href}
\end{align}
where we have introduced the compact notation $\epsilon^{ab\ldots}_{ij\ldots}$ for single-particle energy differences.

Starting from a Hartree-Fock Slater determinant $\ket{\Phi}$, we can write the many-body wave function through first order as 
\begin{align}
    \ket{\Psi} &= \ket{\Phi} - \frac{1}{4}\,g\,\sum_{abij}\frac{H_{abij}}{\epsilon^{ab}_{ij}}\nord{\aaO_a\aaO_b\aO_j\aO_i}\ket{\Phi}\notag\\
    &\quad + O(g^2)\,.
\end{align}
Since $H$ is Hermitian, $H_{ijab} = \overline{H_{abij}}$ and the energy through second order can be written as
\begin{equation}
    E = E_\text{ref} - \frac{1}{4}\,g^2\,\sum_{abij}\frac{|H_{abij}|^2}{\epsilon^{ab}_{ij}} + O(g^3)\label{eq:E2_pert}\,.
\end{equation}

\subsection{General Observables}
For observables other than the energy, we consider the perturbative expansion of the expectation value
\begin{equation}
    \expect{O} = \frac{\matrixe{\Psi}{O}{\Psi}}{\braket{\Psi}{\Psi}}\,.\label{eq:expect_O}
\end{equation}
Treating the denominator first, we obtain
\begin{equation}
    \frac{1}{\braket{\Psi}{\Psi}} = 1 - g^2\braket{\Psi^{(1)}}{\Psi^{(1)}} + O(g^4)\,,
\end{equation}
hence
\begin{align}
    \expect{O}&= \matrixe{\Psi^{(0)}}{O}{\Psi^{(0)}} \notag\\
    &\quad
        + g \left(\matrixe{\Psi^{(1)}}{O}{\Psi^{(0)}}
        +  \matrixe{\Psi^{(0)}}{O}{\Psi^{(1)}}\right)\notag\\
    &\quad   + g^2 \left(\matrixe{\Psi^{(1)}}{O}{\Psi^{(1)}}
        - \matrixe{\Psi^{(0)}}{O}{\Psi^{(0)}}\braket{\Psi^{(1)}}{\Psi^{(1)}}\right.\notag\\
    &\qquad\qquad \left.
        + \matrixe{\Psi^{(2)}}{O}{\Psi^{(0)}}
        + \matrixe{\Psi^{(0)}}{O}{\Psi^{(2)}}\right)
        \notag\\
    &\quad
        +O(g^3)\,.\label{eq:operator_expansion}
\end{align}

Let us now plug in the operator $O$ in normal ordered form:
\begin{equation}
    O = O_0 + O_1 + O_2\,,
\end{equation}
where the subscript again indicates the particle rank of the operator. Then we have
\begin{equation}
    O_0 = \matrixe{\Psi^{(0)}}{O}{\Psi^{(0)}}
\end{equation}
because of Eq.~\eqref{eq:expect_nord} and $\braket{\Psi^{(0)}}{\Psi^{(0)}}=1$. Since the perturbation is a two-body operator, we also note that
\begin{align}
    \matrixe{\Psi^{(1)}}{O}{\Psi^{(0)}} &= \matrixe{\Psi^{(1)}}{O_2}{\Psi^{(0)}} 
    =\overline{\matrixe{\Psi^{(0)}}{O}{\Psi^{(1)}}}\,,\\
    \matrixe{\Psi^{(2)}}{O}{\Psi^{(0)}} &= \matrixe{\Psi^{(2)}}{O_2}{\Psi^{(0)}}
    =\overline{\matrixe{\Psi^{(0)}}{O}{\Psi^{(2)}}}\,,\\
    \matrixe{\Psi^{(1)}}{O}{\Psi^{(1)}} &=
    O_0 \braket{\Psi^{(1)}}{\Psi^{(1)}}
    + \matrixe{\Psi^{(1)}}{O_1}{\Psi^{(1)}} \notag\\
    &\quad + \matrixe{\Psi^{(1)}}{O_2}{\Psi^{(1)}} \,.
\end{align}
Thus, in Eq.~\eqref{eq:operator_expansion} the $O(g^2)$ terms proportional to $O_0$ cancel, the one-body operator $O_1$ appears at second order, and $O_2$ at first order. 

Evaluating the individual terms according to the rules discussed above, we find that
perturbative corrections through $O(g^2)$ are
\begin{align}
    O^{(2)}_1 &= \matrixe{\Psi^{(1)}}{O_1}{\Psi^{(1)}}\notag\\
    &=
    \frac{1}{2}\sum_{abcij}\frac{H_{ijab} O_{ac} H_{cbij}}{\epsilon^{ab}_{ij}\epsilon^{cb}_{ij}} -
    \frac{1}{2}\sum_{abijk}\frac{H_{abij}O_{ik}H_{kjab}}{\epsilon^{ab}_{ij}\epsilon^{ab}_{kj}}\,,\label{eq:dr_1b_2nd}
\\
    O^{(1)}_2 &= \matrixe{\Psi^{(1)}}{O_2}{\Psi^{(0)}} + \matrixe{\Psi^{(0)}}{O_2}{\Psi^{(1)}}\notag\\
    &=-\frac{1}{4}\sum_{abij}\frac{O_{ijab}H_{abij} + H_{ijab}O_{abij}}{\epsilon^{ab}_{ij}}\,,\label{eq:dr_2b_1st}
\intertext{and}
    O^{(2)}_2 &= \matrixe{\Psi^{(1)}}{O_2}{\Psi^{(1)}} + \matrixe{\Psi^{(0)}}{O_2}{\Psi^{(2)}}\notag\\
    &\qquad + \matrixe{\Psi^{(2)}}{O_2}{\Psi^{(0)}} \notag\\
    &= \frac{1}{8}\sum_{abcdij}\frac{H_{ijab} O_{abcd} H_{cdij}}{\epsilon^{ab}_{ij}\epsilon^{cd}_{ij}}\notag\\
    &\quad +\frac{1}{8}\sum_{abijkl}\frac{H_{abkl} O_{klij} H_{ijab}}{\epsilon^{ab}_{ij}\epsilon^{ab}_{kl}}\notag\\
    &\quad - \sum_{abcijk}\frac{H_{ijab}O_{kbic} H_{ackj}}{\epsilon^{ab}_{ij}\epsilon^{ac}_{kj}}
    + \left(O \leftrightarrow H \right)\label{eq:dr_2b_2nd}
\,,
\end{align}
where the short-hand in the last line indicates permutations of the first three terms in which $O$ is swapped with one of the Hamiltonian operators $H$.

We can readily verify that the application of this approach to the Hamiltonian will yield the second-order energy from the previous section: Since $O_1 = \Href - \Eref$, we find after some index manipulation that
\begin{equation}
    H^{(2)}_1 =  \frac{1}{2}\sum_{abij}\frac{|H_{abij}|^2}{(\epsilon^{ab}_{ij})^2} \left(\epsilon_a -
    \epsilon_i \right)  = \frac{1}{4}\sum_{abij}\frac{|H_{abij}|^2}{(\epsilon^{ab}_{ij})^2} \epsilon^{ab}_{ij}\,,
\end{equation}
and with $O_2 = g\Hpert$ the leading two-body term becomes
\begin{equation}
    H^{(2)}_2 = -\frac{1}{2}\sum_{abij}\frac{|H_{abij}|^2}{\epsilon^{ab}_{ij}}\,.
\end{equation}
Plugging everything into Eq.~\eqref{eq:operator_expansion}, we obtain the second-order energy \eqref{eq:E2_pert}.

We conclude this section by emphasizing an important difference between
the perturbative treatment of the Hamiltonian and that of a general observable.
The normal-ordered zero- and one-body parts of $H$ are treated exactly because 
the working basis for the expansion are the eigenfunctions of $\Href$ (cf.~Eq.~\eqref{eq:eig_Href}),
and only the two-body part of $H$ is treated perturbatively. This is not the
case for a general observable $O$. In essence, we have to
view \eqref{eq:expect_O} as a sum of separate expansions for the different 
normal-ordered contributions of $O$, each of which might have their own 
order-by-order convergence behavior. As we will see in a concrete application 
below, the second-order contribution to $\expect{O_1}$ can be greater than the 
first-order contribution to $\expect{O_2}$, and therefore the dominant 
perturbative correction to the expectation value \eqref{eq:operator_expansion}. 

\section{Modeling Perturbative Correction Terms as Inner Products}\label{sec:JLapplication}
This section provides a framework for modeling perturbative corrections to nuclear ground-state energies and observables as the sum of multiple inner products between tensors, so that each inner product can be approximated according to the geometry-preserving property of JLEs as outlined in Lemma \ref{lem:BasicVecInnProdJL}. The idea is to calculate the inner product of tensors with reduced dimensions to obtain an approximate value of the exact results. In doing so, it is assumed that the data lie on a low-rank Hilbert space of tensors.

\subsection{Second-Order Energy Correction} \label{sec:E^2}
For the sake of brevity, we define and use the following real-valued tensor from the energy denominators in
the perturbative expressions: 
\begin{align}
    D_{pqrs}:= 
        \begin{cases}
        \frac{1}{\epsilon_p + \epsilon_q - \epsilon_r - \epsilon_s} & \text{if}\; n_p=n_q =0\;\text{and}\;  \\
        & n_r=n_s=1\,,\\
        0 & \text{else}\,.
    \end{cases}
    \label{eq:D}
\end{align}
Thus, the entries of the tensor $D$ vanish unless the first two indices refer to particle states, and
the last two to hole states. It will allow us to extend sums that are restricted to either class of
states to the entire single-particle basis in the following, e.g.,
\begin{equation}
    \sum_{abij}\frac{H_{ijab}H_{abij}}{\epsilon^{ab}_{ij}} = \sum_{pqrs}H_{rspq}D_{pqrs}H_{pqrs}\,.
\end{equation}

In the following, we assume spherical symmetry for the nucleus and adopt an 
angular-momentum coupled representation, the so-called $J$-scheme. Then the 
indices $p$ represent groups of energetically degenerate single-particle levels
that are characterized by a tuple of radial, orbital angular momentum, angular momentum, 
and isospin quantum numbers\footnote{We denote the radial quantum number with a
$\nu$ instead of the usual $n$ to avoid confusion with the occupation numbers.}:
\begin{equation}
    p = (\nu_p, l_p, j_p, \tau_p)\,.
\end{equation}
As a consequence, the two-body tensors will have a block structure that we
indicate by using the total angular momentum quantum $J$, e.g., $H^{J}_{pqrs}$.
Physically allowed entries of the tensors must satisfy the conditions
\begin{align}\label{eq:ang_mom_rule}
    |j_p - j_q | \leq J \leq j_p + j_q\,, \quad
    |j_r - j_s | \leq J \leq j_r + j_s\,.
\end{align} 
In the uncoupled representation, we would have $2J+1$ copies of each reduced
tensor $H^J_{pqrs}$, and accordingly, these multiplicities will appear as 
explicit factors in subsequent expressions. 

In the $J$-scheme, the second-order energy correction term is defined as:
\begin{equation}
	\begin{split}
		E^{(2)}=\sum\limits_{J=0}^{J_\text{max}}E^{(2)}\left( J \right),
	\end{split}
	\label{equ:E2_total}
\end{equation}
\noindent where $J_\text{max}$ is the largest total angular momentum that can be obtained by coupling the single-particle angular momenta, and
\begin{align}
		E^{(2)}\left( J \right)&=-\frac{1}{4}\left( 2J+1 \right)\sum\limits_{abij}~H^J_{abij}D^J_{abij}H^J_{ijab}\notag\\
		&=-\frac{1}{4}\left( 2J+1 \right)\langle \widetilde{H}^J,H^J \rangle.
	\label{equ:E2}
\end{align}
Here, we have written the energy correction in terms of the inner product \eqref{eq:def_inner_prod} to set the stage for applying the JL lemma, and we have introduced $\widetilde{H}^J$ as an element-wise product of $D^J$ with $H^J$, i.e.,
\begin{equation}
	\begin{split}
		\widetilde{H}^J_{pqrs}=D^J_{pqrs}H^J_{pqrs}.
	\end{split}
	\label{eq:h_tilde}
\end{equation}

An approximation of Eq.~\eqref{equ:E2} can be computed by randomly projecting $H^J$ and $\widetilde{H}^J$ onto a lower-dimensional space using mode-wise JL embeddings:

\begin{equation}
	\begin{split}
		\widehat{H}^J= H^J\bigtimes_{\ell=1}^{4} \mathbf{A}^{(\ell)},
	\end{split}
	\label{proj_H}
\end{equation}

\noindent and

\begin{equation}
	\begin{split}
		\widehat{\widetilde{H}}^J= \widetilde{H}^J\bigtimes_{\ell=1}^{4} \mathbf{A}^{(\ell)},
	\end{split}
	\label{proj_H_tilde}
\end{equation}

\noindent where $\mathbf{A}^{(k)} \in \mathbb{R}^{m_k \times N}$ are JL matrices and $m_k\leq N$ for $k \in [4]$. Now, with high probability,

\begin{equation}
	\begin{split}
		\langle \widetilde{H}^J,H^J \rangle \approx \langle \widehat{\widetilde{H}}^J,\widehat{H}^J \rangle,
	\end{split}
\end{equation}

\noindent to within an adjustable error that is related to the target dimension sizes $m_j$.

A second-stage JL embedding can be applied to the vectorized versions of $H^J$ and $\widetilde{H}^J$ to further compress the projected tensors before computing the approximate inner product. This is achieved by calculating
\begin{equation}
	\widehat{\mathbf{h}} = \mathbf{A} \operatorname{vect}\left(\widehat{H}^J\right),
\end{equation}

\noindent where $\widehat{\mathbf{h}} \in \mathbb{R}^m$ and $\mathbf{A} \in \mathbb{R}^{m \times \prod_{k=1}^{4}m_k}$. The same operation is performed on $\widetilde{H}^J$ to obtain $\widehat{\widetilde{\mathbf{h}}}$.

\subsection{Radius Corrections}
To explore the impact of the JL embeddings on observables, we consider
the mean-square radius operator\footnote{We suppress the exponent of the operator to reduce clutter in subsequent expressions.}
\begin{equation}
    R := \frac{1}{A}\sum_{i=1}^A \left(\vec{r}_i - \vec{r}_\text{cm}\right)^2\,,
\end{equation}
with 
\begin{equation}
    \vec{r}_{cm} = \frac{1}{A}\sum_{i=1}^A\vec{r}_i\,.
\end{equation}
We write it as a sum of one- and two-body operators,
\begin{align}\label{eq:def_R}
    R &= \frac{1}{A}\left(\left(1-\frac{1}{A}\right)\sum_{i}\vec{r}_i^{\,2} - \frac{1}{A}\sum_{i\neq j}\vec{r}_i\cdot\vec{r}_j\right)\,,
\end{align}
and normal order it with respect to the Hartree-Fock reference state $\ket{\Phi}$ before plugging it into Eqs.~\eqref{eq:dr_1b_2nd} and \eqref{eq:dr_2b_1st} to evaluate the leading radius corrections. In the following, we discuss how the individual contributions can be written as inner products along the lines of Eq.~\eqref{equ:E2}.

\subsubsection{One-Body Operator, Particle Term}
The first one-body correction term in Eq.~\eqref{eq:dr_1b_2nd} contains
a sum over the particle matrix elements $R_{ac}$ of the normal-ordered one-body part of the radius operator. It can be expressed in the following way:
\begin{align}
    R_I &= \frac{1}{2}\sum_{J}(2J+1)\sum_{abcij}H^J_{ijab}D_{abij} R_{ac} D_{cbij} H^J_{cbij}\notag\\
    &= \frac{1}{2}\sum_{J}(2J+1)\sum_{abcij}\overline{H^J_{abij}}D_{abij} R_{ac} D_{cbij} H^J_{cbij}\notag\\
    &= \frac{1}{2}\sum_{J}(2J+1)\sum_{abcij}\overline{\widetilde{H}^J_{abij}} R_{ac} \widetilde{H}^J_{cbij}\notag\\
    &= \frac{1}{2}\sum_{J}(2J+1)\sum_{abij}\overline{\widetilde{H}^J_{abij}} X^J_{abij}\notag\\
    &=\frac{1}{2}\sum_J (2J+1)\left\langle X^J, \widetilde{H}^J \right\rangle,\label{R_1}
\end{align}

\noindent where $\widetilde{H}$ is defined as in Eq.~\eqref{eq:h_tilde}, and we have used that $D$ is
real-valued, and $H$ is Hermitian. Note that the relation 
\begin{equation}
    X^J_{pqrs}=\sum_a R_{pa}\widetilde{H}_{aqrs}^J  =\sum_t R_{pt} \widetilde{H}_{tqrs}^J .
    \label{eq:X}
\end{equation}
holds for the newly introduced tensor $X$ due to the properties of $D$ (cf.~Eq.~\eqref{eq:D}).
We also observe that 
\begin{equation}
    X^J=\widetilde{H}^J\times_1 \mathbf{R},
\end{equation}
where $\mathbf{R}$ is the matrix of coefficients of the one-body part of the normal
ordered radius operator. 
Combining everything, the approximate particle term can be calculated as
\begin{equation}
    R_I\approx \frac{1}{2} \sum_{J}(2J+1) \langle \widehat{X}^J, \widehat{\widetilde{H}}^J \rangle,
\end{equation}
where
\begin{equation}
    \widehat{\widetilde{H}}^J=\widetilde{H}^J \bigtimes_{\ell=1}^{4} \mathbf{A}^{(\ell)},
    \label{eq:H_p1}
\end{equation}
and
\begin{equation}
    \widehat{X}^J=X^J \bigtimes_{\ell=1}^{4} \mathbf{A}^{(\ell)}=\widetilde{H}^J\times_1\left(\mathbf{A}^{(1)}\mathbf{R}\right)\bigtimes_{\ell=2}^4 \mathbf{A}^{(\ell)}.
\end{equation}

\subsubsection{One-Body Operator, Hole Term}
The calculation of the second term is very similar to that for the first term  --- we do
need to be mindful that the summation now involves the hole matrix elements $R_{ik}$:
\begin{align}
        R_{II} 
        &=\frac{1}{2}\sum_J(2J+1)\sum\limits_{abijk}H^J_{abij}D_{abij}H^J_{kjab}D_{abkj}R_{ik}\notag\\
        &=\frac{1}{2}\sum_J(2J+1)\sum\limits_{abijk}\widetilde{H}^{J}_{abij}R_{ik}\overline{\widetilde{H}^J_{abkj}}\notag\\
        &=\frac{1}{2}\sum_J(2J+1)\sum\limits_{abij}Y^{J}_{abij}\overline{\widetilde{H}^J_{abij}}\notag\\
        &=\frac{1}{2}\sum_J(2J+1)\left\langle Y^J, \widetilde{H}^J \right\rangle,
    \label{R_2}
\end{align}

\noindent where we have introduced
\begin{equation}
    Y^J_{pqrs}=\sum_i \widetilde{H}^J_{pqis}R_{ir} = \sum_t \widetilde{H}^J_{pqts}R_{tr}.
\end{equation}
We can again rely on the properties of $\widetilde{H}^J$ and $D$ to write the sum over holes 
as an unrestricted sum over the single-particle indices.
Since
\begin{equation}
 Y^J=\widetilde{H}^J\times_3 \mathbf{R}^\top,
\end{equation}
we can define
\begin{align}
    \widehat{Y}^J
    &=Y^J \bigtimes_{\ell=1}^{4} \mathbf{A}^{(\ell)}\notag\\
    &=\widetilde{H}^J\times_1\mathbf{A}^{(1)} \times_2\mathbf{A}^{(2)} \times_3 (\mathbf{A}^{(3)}\mathbf{R}^\top)\times_4\mathbf{A}^{(4)}
\end{align}
and write the approximate hole term as
\begin{equation}
  R_{II}\approx \frac{1}{2}\sum_{J}(2J+1)\langle \widehat{Y}^J, \widehat{\widetilde{H}}^J \rangle.  
\end{equation}

\subsubsection{Two-Body Operator}
\label{sec:radius_twobody}
The leading radius correction from the normal-ordered two-body operator, Eq.~\eqref{eq:dr_2b_1st}, has the same structure as the second-order energy correction. Using the Hermiticity of $H$ and $R$, we have
\begin{align}
    R^{(1)}_2
    &=-\frac{1}{2}\sum_{J}(2J+1)\sum_{abij}\Re R^J_{abij}\widetilde{H}^J_{abij}\notag\\
    &=-\frac{1}{2}\sum_{J}(2J+1)\Re\;\langle R^J, \widetilde{H}^J  \rangle\,,\label{eq:R_2B_LO}
\end{align}
and after projection,
\begin{align}
    R^{(1)}_2
    &\approx-\frac{1}{2}\sum_{J}(2J+1)\Re\;\langle \widehat{R}^J, \widehat{\widetilde{H}}^J  \rangle\,.\label{eq:R_2B_LO_approx}
\end{align}

\section{Applications}
\label{sec:JLresults}
In this section, numerical results are provided to demonstrate how modewise JLEs affect the accuracy of MBPT calculations.

\subsection{Preliminaries}

\subsubsection{Interactions and Single-Particle Bases}
\label{sec:bases}
We perform Hartree-Fock and MBPT(2) calculations in spherical harmonic oscillator (HO) working bases that are characterized by the oscillator frequency $\hbar\omega$ and the energy quantum number $e=2\nu+l$, where $\nu$ and $l$ are the oscillator's radial and orbital angular momentum quantum numbers, respectively. The basis is truncated by imposing $e\leq \eMax$. For $\eMax>10$, we introduce an additional truncation $l\leq \lMax = 10$ that further reduces the single-particle basis dimension. Table \ref{tab:basis_size} summarizes these basis dimensions --- and therefore the mode dimensions --- for the different values of $\eMax$ we consider in the following (also see Sec.~\ref{sec:JLapplication}).
For a given $\eMax$ and $\lMax$, this the total angular momentum $J$ is restricted by
\begin{equation}
    0 \leq J \leq 2\;\min\left(\eMax, \lMax\right)  + 1\,.
\end{equation}
\begin{table}[t]
    \centering
    \begin{tabular}{|l|*{6}l|}
        \hline
         $\eMax$ & 4 & 6 & 8 & 10 & $12^*$ & $14^*$   \\
         \hline
         dimension & 30 & 56 & 90 & 132 & 174 & 216 \\
        \hline
    \end{tabular}
    \caption{Basis truncation parameters and mode dimensions for single-particle bases labeled by $\eMax$ in the following. For $\eMax=12,14$ we introduce an additional truncation on the single-particle orbital angular momentum, $l\leq \lMax=10$. See text for details.}
    \label{tab:basis_size}
\end{table}

Our starting Hamiltonians consist of two- and three-nucleon interactions from Chiral Effective Field Theory. In particular, we will consider the family of interactions introduced in \cite{Hebeler:2011dq,Nogga:2004il}, which we label as EM$\lambda/\Lambda$ for short in the following. They consist of the chiral N${}^3$LO nucleon-nucleon interaction by Entem and Machleidt, whose resolution scale $\lambda$ has been lowered by Similarity Renormalization Group evolution \cite{Bogner:2010pq}, and an N${}^2$LO three-nucleon interaction with momentum cutoff $\Lambda$ whose parameters have been adjusted to fit the binding energy and charge radius of $\nuc{He}{4}$. 

We handle the enormous memory requirements of the three-nucleon interaction in the usual way, by introducing a truncation on the energy of (HO) three-nucleon states:
\begin{equation}\label{eq:E3max}
    e_1 + e_2 + e_3 \leq \min( 3\eMax, \EMax)\,,\quad \EMax=14\,.
\end{equation}

The solution of the Hartree-Fock equations is used as a reference state to normal order the Hamiltonian and other observables (cf.~Sec.~\ref{sec:no}). We switch to the HF single-particle basis and employ the commonly used normal-ordered two-body approximation (NO2B) (cf. Sec.~\ref{sec:no} and Refs.~\cite{Roth:2012qf,Hergert:2016jk,Tichai:2020ft}) and discard the residual normal ordered three-nucleon operators in the evaluation of MBPT corrections. In the medium-mass nuclei we are considering in the following, this causes a systematic error of about 1-2\% in the energy and other observables \cite{Roth:2012qf,Gebrerufael:2016fe}. We note that this error has no impact on the performance of the JLEs, which are the focus of the present study. In follow-up work, we will explore whether JLE-based compression can make the explicit inclusion of the truncated three-nucleon terms in MBPT and other types of many-body calculations computationally feasible.

\subsubsection{Compression and Error Measures}
\label{sec:measures}
For a general tensor, we can define the compression in mode $k$ as
\begin{equation}
	c_k=\frac{m_k}{n_k},
	\label{equ:comp}
\end{equation}
\noindent where $n_k$ and $m_k$ denote the size of mode $k$ before and after projection, respectively. The target dimension $m_k$ in JL matrices is chosen as $m_k=\left \lceil c_k n_k\right \rceil$ for all $k$ to ensure that at least a fraction $c_k$ of the ambient dimension in mode $k$ is preserved. 

The focus of the present work is on the coefficient tensor $H_{pqrs}$ of the normal-ordered Hamiltonian (see Sec.~\ref{sec:no}), since it drives the computational and storage costs of the MBPT(2) method. This tensor is not only Hermitian, but also antisymmetric under permutations $p\leftrightarrow q$ and $r \leftrightarrow s$ (cf.~Eq.~\eqref{eq:perm}). Thus, all of its modes have the same dimension, and we choose the same compression for all modes, i.e.,
\begin{equation}
    n_k = N\;, \quad c_k = c\;, \quad  \text{for all}\; k.
\end{equation}
The total compression, denoted by $c_{tot}$ in the following, is defined by the number of elements in a tensor after compression divided by the number of elements in the uncompressed version. For single-stage JL embeddings, it is given by
\begin{equation}
    \ctot=\prod_{k=1}^d\frac{ m_k}{n_k}=\prod_{k=1}^d \frac{\left \lceil c_k n_k\right \rceil}{n_k}
    =\frac{\left \lceil c N\right \rceil ^4}{N^4}
\end{equation}
and for two-stage embeddings (cf.~Sec.~\ref{sec:E^2}), we have
\begin{equation}
    \ctot =\frac{\left\lceil c_2 \prod_{k=1}^d m_k \right\rceil}{\prod_{k=1}^d n_k}
    =\frac{\left \lceil c_2 \left \lceil c N\right \rceil^4 \right \rceil}{N^4}\,.
\end{equation}
This total compression is applied uniformly in all $J$ channels in the present study.

Next, we introduce the error measures we will be using to assess the performance of the JLEs in the following. For specific perturbative contributions, it is natural to consider the mean error
\begin{equation}
	\overline{\Delta O}={\rm mean}\left(~\left| \widehat{O} - O \right|~\right),
	\label{eq:error_general}
\end{equation}
where $\widehat{O}$ is evaluated using the projected and compressed inner products defined 
in Sec.~\ref{sec:JLapplication}, and we take the mean over a number of trials (typically 100 or more).

In a broader context, the measure \eqref{eq:error_general} does not account for the fact that the quantities we are approximating with JLEs are mere corrections to the observables were are interested in, e.g., the total ground-state energy and mean-square radius of a nucleus. For that reason, in what follows, we will also consider 
\begin{align}
	\overline{\Delta E}&={\rm mean}\left(~\left| \widehat{E} - E \right|~\right),
	\label{equ:error_E_mae}
\end{align}
where $E=E_0 + E^{(2)}=\Eref + E^{(2)}$, and
\begin{align}
	\overline{\Delta R}&={\rm mean}\left(~\left| \widehat{R} - R \right|~\right),
	\label{equ:error_R_mae}
\end{align}
where $R=R_0 + R_I + R_{II}$ (see Sec.~\ref{sec:radius_corrections}).

\subsection{Choice of Johnson-Lindenstrauss Embedding}
\label{sec:JLscheme}
\begin{figure*}[t]
    \centering
    \begin{subfigure}{0.46\textwidth}
        \centering
        \includegraphics[width=0.9\linewidth]{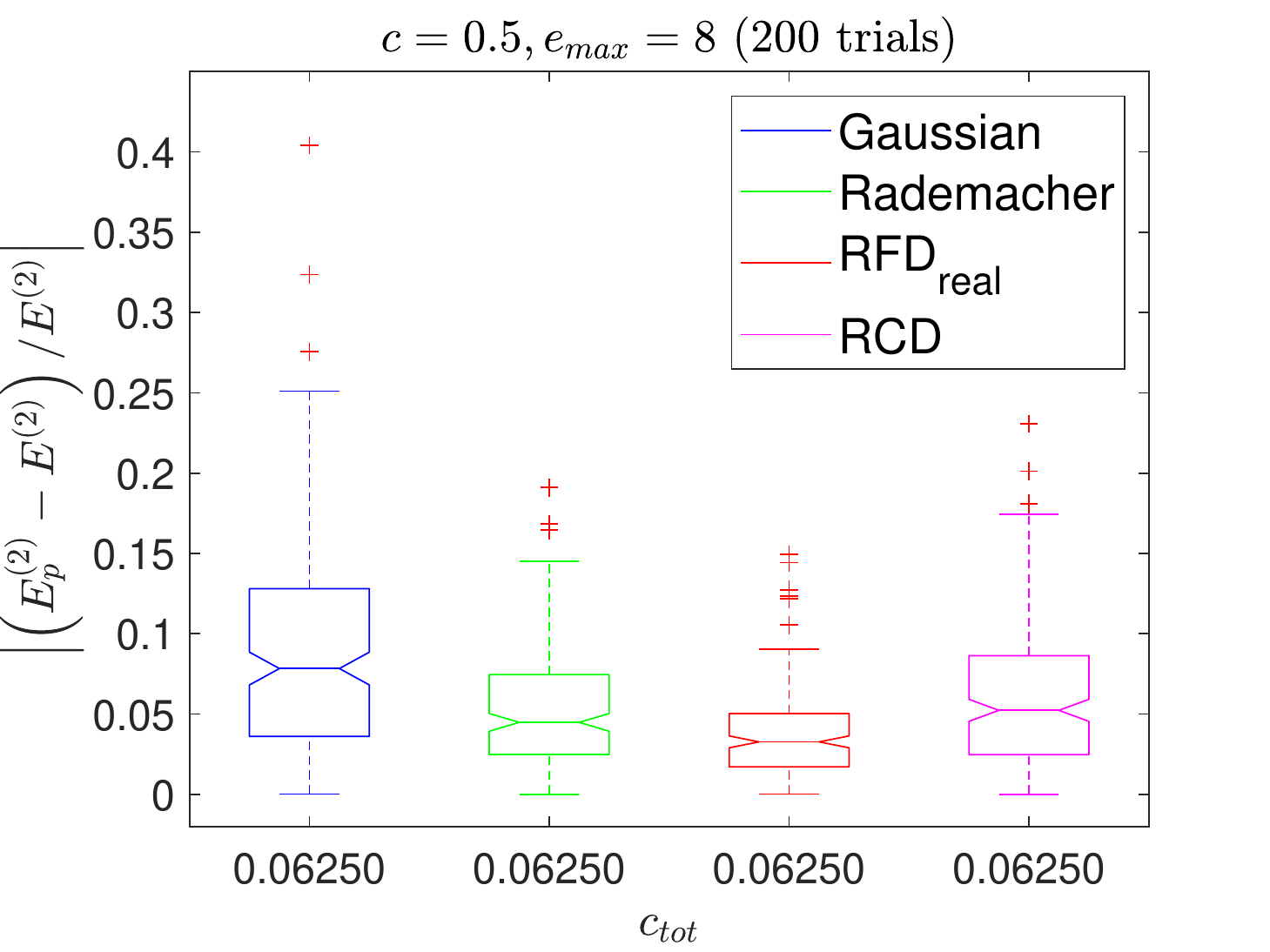}
        \caption{}
    \end{subfigure}
    \begin{subfigure}{0.46\textwidth}
        \centering
        \includegraphics[width=0.9\linewidth]{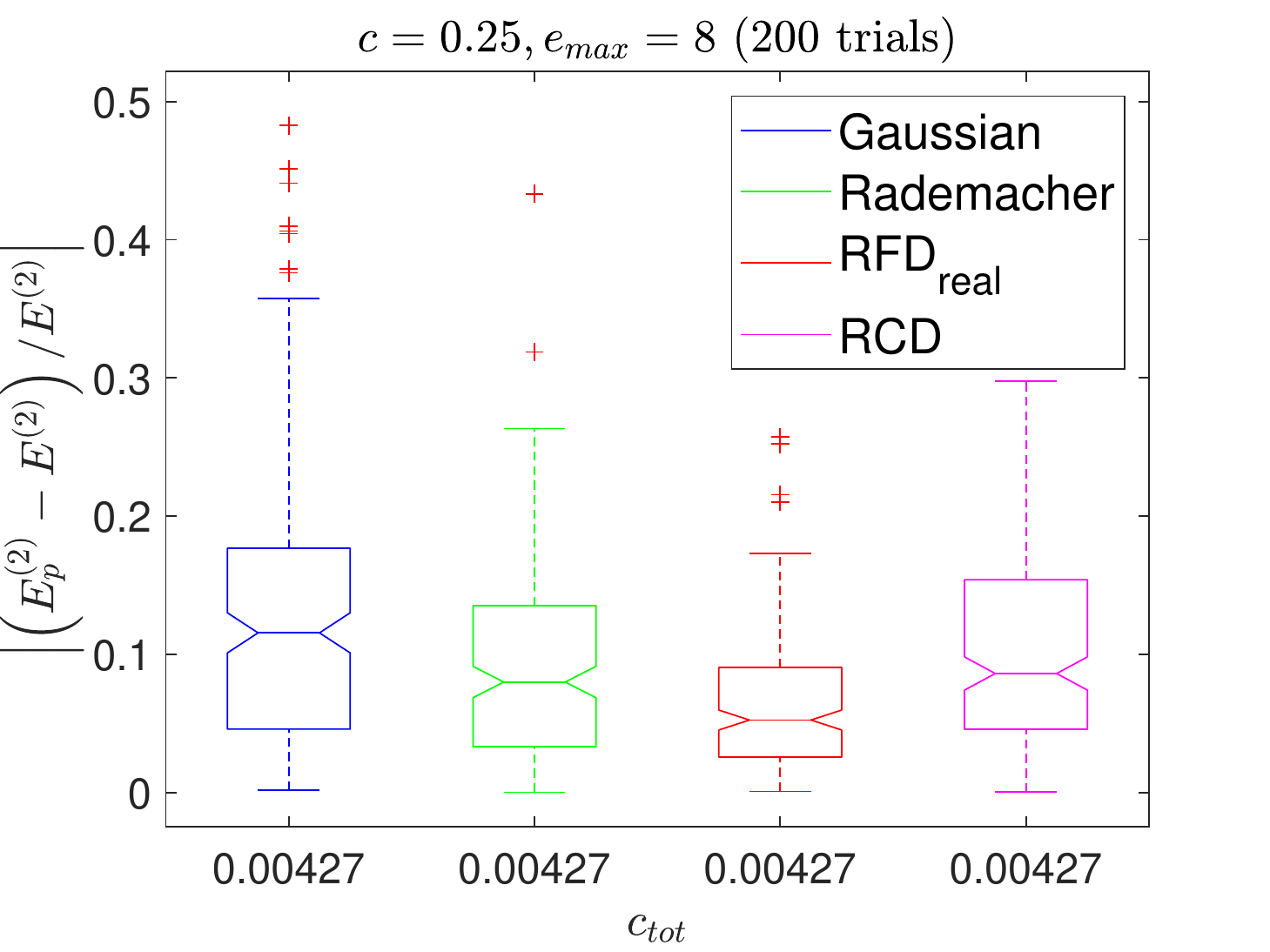}
        \caption{}
    \end{subfigure} \begin{subfigure}{0.46\textwidth}
        \centering
        \includegraphics[width=0.9\linewidth]{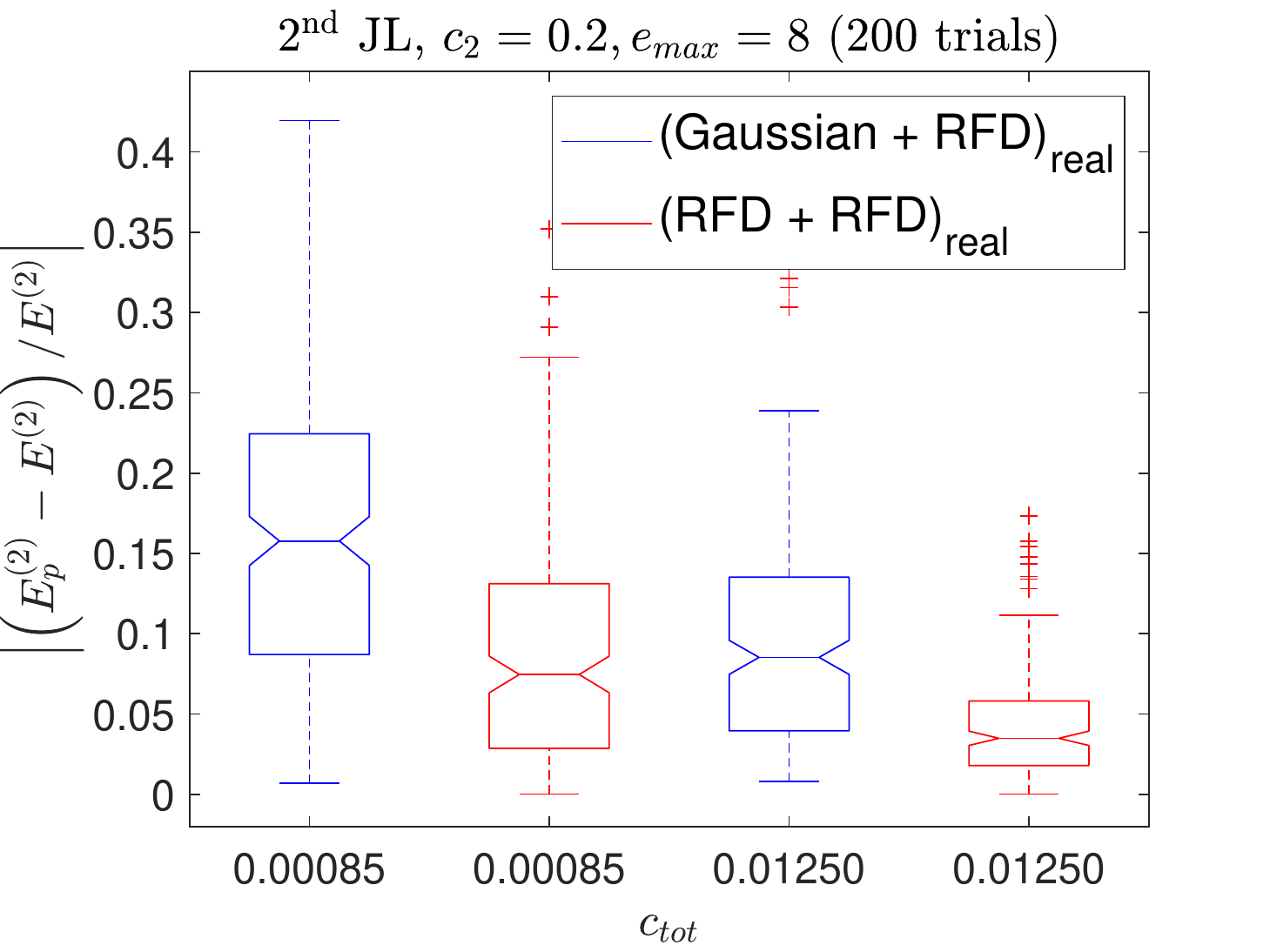}
        \caption{}
    \end{subfigure}
    
    \caption{Performance of different one- and two-stage JL schemes for evaluating  $E^{(2)}$ in $\nuc{O}{16}$. In the box plots, the central mark indicates the median, and the bottom and top edges of the box indicate the $25^{\rm th}$ and $75^{\rm th}$ percentiles, respectively. The red cross symbols show the outliers. 
    Panel (c) shows the impact of applying a second RFD stage of compression to the Gaussian and RFD results shown in panels (a) and (b) --- see text for details. Input data were generated for the EM$1.8/2.0$ interaction with $\eMax=8,\hbar\omega=24\,\MeV$.
    }
    \label{fig:4-method_comparison_box}
\end{figure*}

Many prescriptions exist for selecting the matrices $\mathbf{A}$ (cf. ~Definition \ref{def:jl_matrices}) that are used as JL embeddings. For that reason, we first explore the performance of several common choices, namely Gaussian \cite{dasgupta2003elementary}, Rademacher \cite{achlioptas2003database}, and the so-called Fast JL \cite{krahmer2011new} matrices.
For \emph{Gaussian JL},
\begin{equation}
 \mathbf{A}^{(j)}=\frac{1}{\sqrt{m_j}}\mathbf{G}   
\end{equation}
is used for all $j \in [d]$ where $d$ is the number of modes in the tensors forming the inner product, $m_j$ is the target dimension for mode $j$, and each entry in $\mathbf{G}$ is an independent and identically distributed standard Gaussian random variable $\mathbf{G}_{ij}\sim \mathcal{N}\left( 0,1 \right)$.
For \emph{Rademacher JL},
\begin{equation}
    \mathbf{A}^{(j)}=\frac{1}{\sqrt{m_j}}\mathbf{A},
\end{equation}
where the elements of $\mathbf{A}$ are Rademacher random variables, i.e., they take on the values $+1$ and $-1$ with equal probabilities.
For \emph{Fast JL}, we will use the definitions
\begin{equation}\label{eq:rfd}
    \mathbf{A}^{(j)}=\frac{1}{\sqrt{m_j}}\mathbf{R}\mathbf{F}\mathbf{D}    
\end{equation}
or 
\begin{equation}\label{eq:rcd}
    \mathbf{A}^{(j)}=\frac{1}{\sqrt{m_j}}\mathbf{R}\mathbf{C}\mathbf{D}
\end{equation}
for all $j \in [d]$, where $\mathbf{R}$ denotes the random restriction matrix which uniformly picks rows from the matrix it is applied to, $\mathbf{F}$ and $\mathbf{C}$ are the unitary Discrete Fourier Transform and type-$1$ Discrete Cosine Transform matrices scaled by $\sqrt{n_j}$, respectively, and $\mathbf{D}$ is a diagonal matrix of Rademacher random variables \cite{krahmer2011new}. The two Fast JL schemes are labeled RFD and RCD in the following. For the RFD scheme, we make the additional simplification of considering only the real part of the result when applying the Fourier matrices --- we indicate this through the subscript $(\cdot)$\textsubscript{real}.

\begin{figure}[t]
	\centering
		\includegraphics[width=0.9\linewidth]{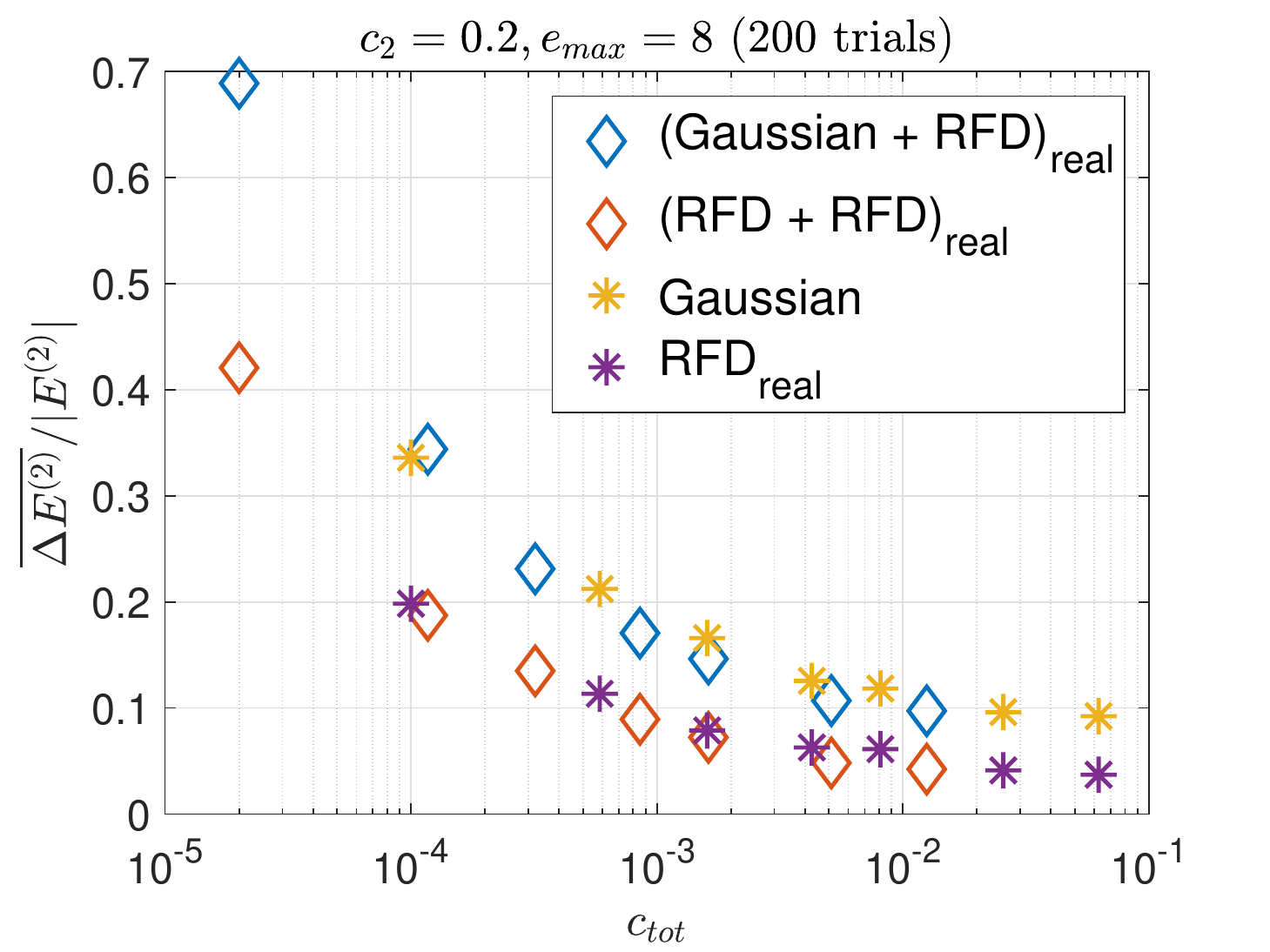}
	\caption{Relative mean error in $E^{(2)}$ in $\nuc{O}{16}$ as a function of the compression $c_{tot}$ for one- and two-state JL schemes (cf.~Fig.~\ref{fig:4-method_comparison_box}). Input data were generated for the EM$1.8/2.0$ interaction with $\eMax=8,\hbar\omega=24\,\MeV$. }
 	\label{fig:4-method_comparison_mean}
\end{figure}

In Figures \ref{fig:4-method_comparison_box} and \ref{fig:4-method_comparison_mean} we compare the performance of the aforementioned JL embeddings for the second-order energy correction in $\nuc{O}{16}$, using a representative chiral two- plus three-nucleon Hamiltonian. From the data shown in panels (a) and (b), it is evident that the RFD$_\mathrm{real}$ scheme emerges as the clear favorite of the four options we are considering here. If we reduce the size of the $\eMax=8$ tensors' single-particle basis by $50\%$, so that the overall compression is $c_\mathrm{tot}=0.5^4 = 0.0625$, the median error of $E^{(2)}$ over 200 trials is about 4\%, and even the outliers do not exceed 15\%. These errors correspond to $\sim 1\%-4\%$ errors for the total MBPT(2) ground-state energy. Applying a more aggressive compression with $c=23/90\approx 0.255$, the median error for the RFD$_\mathrm{real}$ scheme increases only moderately to about 5\%, although the distribution of the errors spreads notably, with outliers reaching more than 20\%. The increase in median error and spread of the distribution are much bigger for the other schemes.

In Fig.~\ref{fig:4-method_comparison_box}(c), we show the effect of applying a second-stage JL embedding with $c_2=0.2$ to the Gaussian and RFD$_\mathrm{real}$ JL results of panels (a) and (b). It is perhaps not surprising that an additional compression causes an increase of the median error and standard deviation of the data. Note, however, that the two-stage (RFD \text{+} RFD)\textsubscript{real} scheme achieves a five-fold compression over the $\ctot=0.0625$ RFD\textsubscript{real} results to $\ctot=0.0125$ while approximately maintaining the width of the error distribution and the median, which increases from $3.26$\% to $3.48$\%. For the total ground-state energy, the (RFD \text{+} RFD)\textsubscript{real} scheme achieves sub-percent errors while only retaining 1.25\% of the coefficients in the two-body part of the Hamiltonian.

In Fig.~\ref{fig:4-method_comparison_mean}, we show the mean error of the one- and two-stage JL schemes used in Fig.~\ref{fig:4-method_comparison_box}(c) as a function of the total compression $\ctot$. Evidently, the two RFD\textsubscript{real} schemes yield the lowest mean errors at a given level of compression. Since they consistently outperform the other JL methods in terms of mean, median and standard deviation of our results, we will focus on these schemes in the following, and we will use (RFD \text{+} RFD)\textsubscript{real}, in particular, because it offers the best compromise between compression and precision.

We conclude here with an observation that is relevant for future research: The mode-wise application of the matrices $\mathbf{F}\mathbf{D}$ or $\mathbf{C}\mathbf{D}$ (cf. Eqs.~\eqref{eq:rfd} and \eqref{eq:rcd}) can essentially be viewed as a change of the single-particle basis in which the tensors are represented, even if the restriction to real values eventually makes the transformation projective instead of unitary. Clearly, this change of basis ahead of the random sampling is beneficial, since it reduces the embedding errors. There are also a variety of physics-inspired approaches for optimizing the single-particle basis. As explained above, the Hartree-Fock basis, which is our starting representation, is obtained by minimizing the ground-state energy for a particular class of reference state, and perturbatively enhanced ``natural'' orbitals have recently been used for compression and convergence acceleration in nuclear many-body theory \cite{Tichai:2019to,Hoppe:2021ij}. In future work, we will explore the interplay and possible integration of such basis optimization techniques with JLEs.

\subsection{Energy Corrections}
\label{sec:energy_corrections}

\subsubsection{General Features}

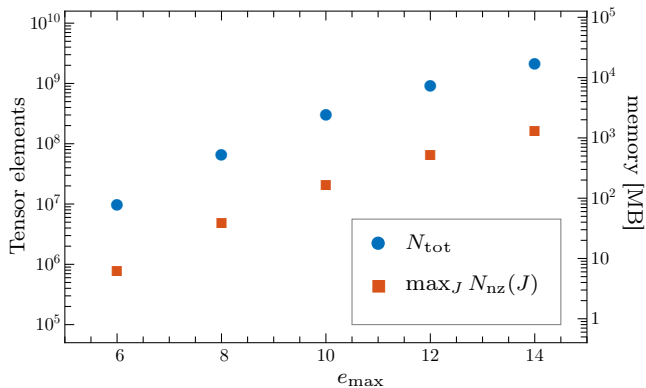
\begin{figure}[t]
    \centering
    \setlength{\unitlength}{\columnwidth}
    \begin{picture}(1.0000,0.6800)
        \put(0.0300,0.0500){\includegraphics[width=0.92\unitlength]{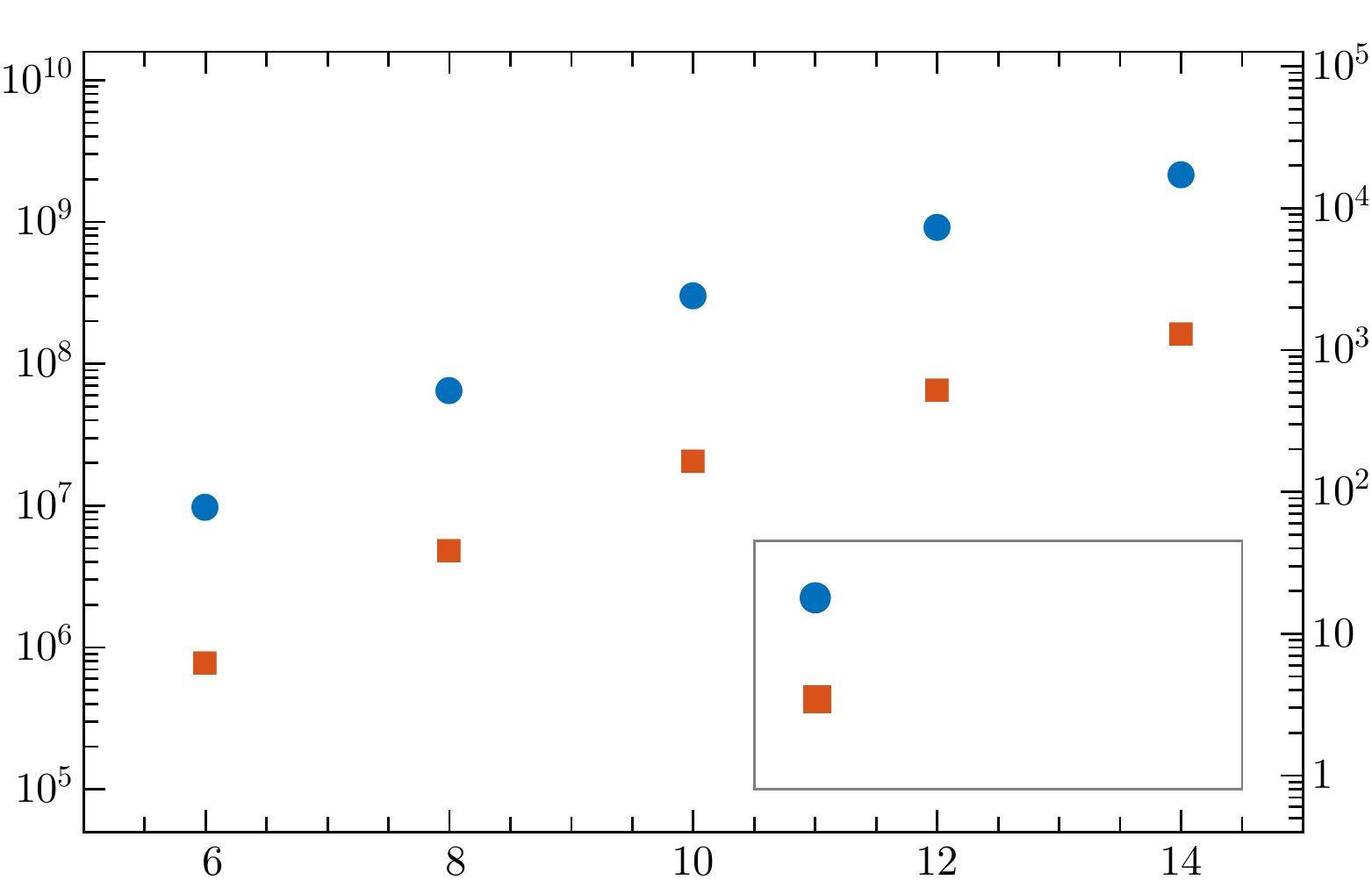}}
        \put(0.0000,0.1000){\rotatebox{90}{\parbox{0.5000\unitlength}{\centering Tensor elements}}}
        \put(0.9600,0.6200){\rotatebox{-90}{\parbox{0.5000\unitlength}{\centering memory [MB]}}}
        \put(0.6200,0.2350){\parbox{0.08\unitlength}{\raggedright $\Ntot$}}
        \put(0.6200,0.1700){\parbox{0.08\unitlength}{\raggedright $\max_J N_\text{nz}(J)$}}
        \put(0.1000,0.0200){\parbox{0.9\unitlength}{\centering$\eMax$}}
    \end{picture}
    \caption{Total number of Hamiltonian tensor elements, $\Ntot$, and number of physically allowed nonzero entries, $\max_J N_\text{nz}(J)$, for each $J$ channel as a function of the basis size $\eMax$. The right axis shows the associated memory required for storage. See text for details.}
    \label{fig:num_nonzero}
\end{figure}

Now that we have identified a favored JL scheme, we will proceed and explore its performance for different nuclei, basis (and tensor) sizes, and interactions. To provide context for the subsequent discussion, we first consider some general features of the Hamiltonian tensor as well as the second-order MBPT corrections. 

Note that a typical single-particle basis without any symmetry restrictions 
can consist of well above a 1000 states, especially if it also must account for weakly bound 
nucleons with spatially extended single-particle wave functions. In this representation, the 
Hamiltonian naively would have $10^{12}$ or more elements, although it can be extremely sparse 
due to the symmetries of the interaction. For particular applications, we can impose symmetries 
like rotational invariance, and achieve more manageable requirements: In the $J$-scheme with 
explicit spherical symmetry (cf. Sec.~\ref{sec:JLapplication}), the tensor becomes block diagonal 
and the blocks with fixed angular momentum $J$ typically range from $\Ntot=10^{7}-10^{10}$ elements, 
as illustrated in Fig.~\ref{fig:num_nonzero}. Physical conservation laws for parity and isospin or 
charge force many of these elements to vanish and allow a reduction by an additional order of 
magnitude. In Fig.~\ref{fig:num_nonzero}, we show the resulting number of nonzero entries 
$N_\text{nz}$ of the largest $J$ channel for each basis size $\eMax$. In a typical application,
we have between 10 and 20 of these channels, and about a quarter of them have comparably large
$N_\text{nz}(J)$, while the remaining channels are very small in comparison. As we can see, this translates
into memory requirements in the 10 MB to 10 GB range\footnote{Depending on the storage format for the
sparse Hamiltonian tensor, we may face significant overhead: A coordinate-based format, for instance, would have a
fivefold overlap because it needs to store four indices for each element in addition to the element's 
value.}. 

\begin{figure}[t]
    \centering
    \setlength{\unitlength}{\columnwidth}
    \begin{picture}(1.0000,0.6800)
        \put(0.0500,0.0500){\includegraphics[width=0.95\unitlength]{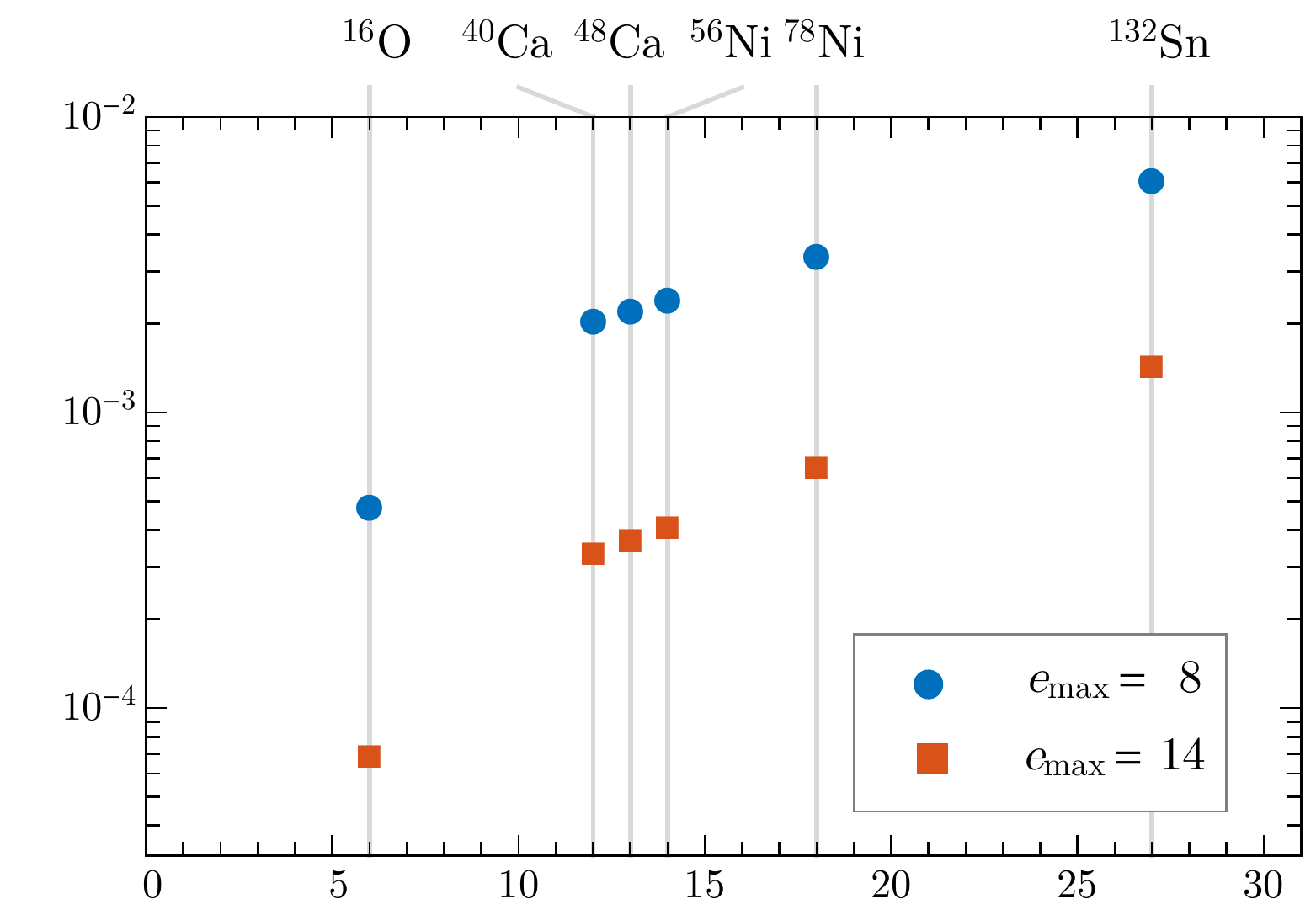}}
        \put(0.0200,0.1000){\rotatebox{90}{\parbox{0.5000\unitlength}{\centering$\max_J N^{(2)}(J)/N_\mathrm{tot}$}}}
        \put(0.1000,0.0200){\parbox{0.9\unitlength}{\centering$N_o$}}
    \end{picture}
    \caption{Maximum fraction of Hamiltonian coefficients per $J$ channel that can contribute to $E^{(2)}$ as a function of the number of occupied ($N_o$) single-particle orbitals. See text for details.}
    \label{fig:num_pphh}
\end{figure}

Switching focus to the many-body method, we recall from Sec.~\ref{sec:mbpt} that the first-order 
wave function correction $\ket{\Psi^{(1)}}$ and second-order energy correction $E^{(2)}$ only depend on a 
subset of elements of the Hamiltonian tensor, namely $H_{abij}$ and $H_{ijab}$. As a reminder $a,b$ 
refer to unoccupied (particle) orbitals, while $i,j$ are occupied (hole) orbitals, and there are much 
fewer of the latter than the former, so that $N_o \ll N_u$ and $N = N_o + N_u$.
The overall number of elements of the Hamiltonian that can contribute to $E^{(2)}$ (and $\ket{\Psi^{(1)}}$) 
will then be given by
\begin{equation}\label{eq:N2}
    N^{(2)} = 2 N_o^2 N_u^2\,,
\end{equation}
and the fraction of overall elements this corresponds to is
\begin{equation}\label{eq:E2_fraction}
    \frac{N^{(2)}}{N_\mathrm{tot}} = \frac{2 N_o^2 N_u^2}{N^4}\,.
\end{equation}

Since we are working in the $J$-scheme, $N_o$ does not directly map to the number of nucleons because 
each orbital can be multiply occupied. Moreover, there is no simple analytical expression that generalizes
\eqref{eq:N2} to each channel because of the angular momentum selection rule \eqref{eq:ang_mom_rule}, 
although $N_\mathrm{tot}=N^4$ continues to hold. Counting the elements explicitly, we obtain the maximal 
fraction $N^{(2)}(J)/N_\mathrm{tot}$ across $J$ channels for the nuclei we will discuss in the following, 
which is shown in Fig.~\ref{fig:num_pphh}.  We see that for calcium and heavier nuclei, 
it lies between 0.1\% and 1\% for $\eMax=8$, and it is a factor 5-6 smaller for $\eMax=14$
because $N_u$ grows with $\eMax$ while $N_o$ stays constant. For the future discussion, we note that
the fraction of relevant tensor elements is substantially smaller for $\nuc{O}{16}$ than for the other
nuclei, as expected because of the smaller number of occupied orbitals.
Note that these numbers should be understood as upper bounds, since the natural energy scales of the 
interaction can limit the size of formally relevant tensor elements.

\begin{figure*}
    \centering
    \begin{subfigure}{0.46\textwidth}
        \centering
        \includegraphics[width=0.9\linewidth]{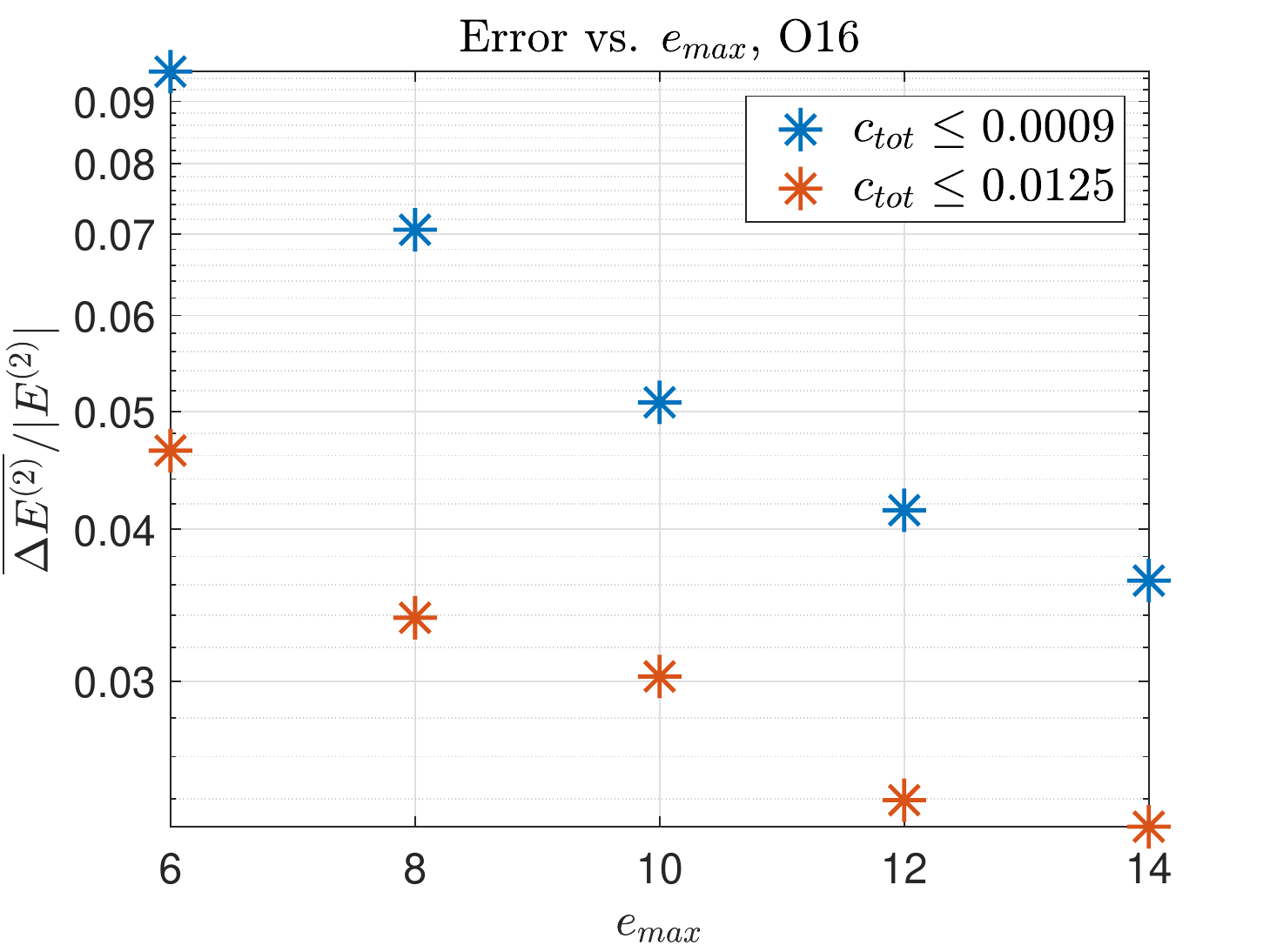}
    \end{subfigure}
    \begin{subfigure}{0.46\textwidth}
        \centering
        \includegraphics[width=0.9\linewidth]{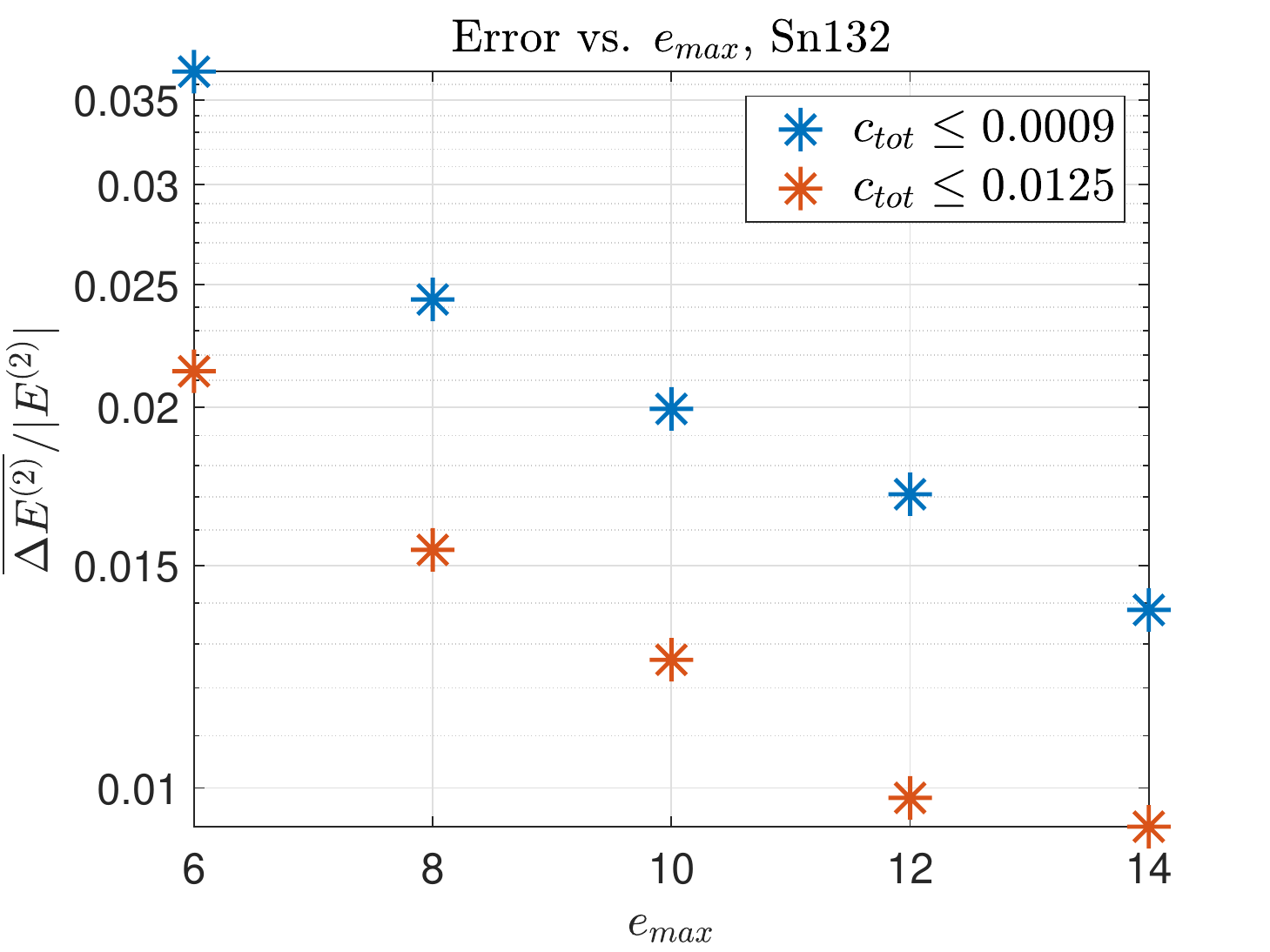}
    \end{subfigure}
    
    \caption{Mean relative error of the second-order energy correction, $|\overline{\Delta E^{(2)}}/E^{(2)}|$, for $\nuc{O}{16}$ and $\nuc{Sn}{132}$ as a function of the basis size $\eMax$. All calculations were performed with the EM$1.8/2.0$ interaction, using the two-stage (RFD \text{+} RFD)\textsubscript{real} JL embedding and $200$ trials.}
    \label{fig:error_vs_emax}
\end{figure*}

\subsubsection{Second-Order Energy Correction}
\label{sec:E2}

\begin{figure*}[t]
    \centering
    \begin{subfigure}{0.46\textwidth}
        \centering
        \includegraphics[width=0.9\linewidth]{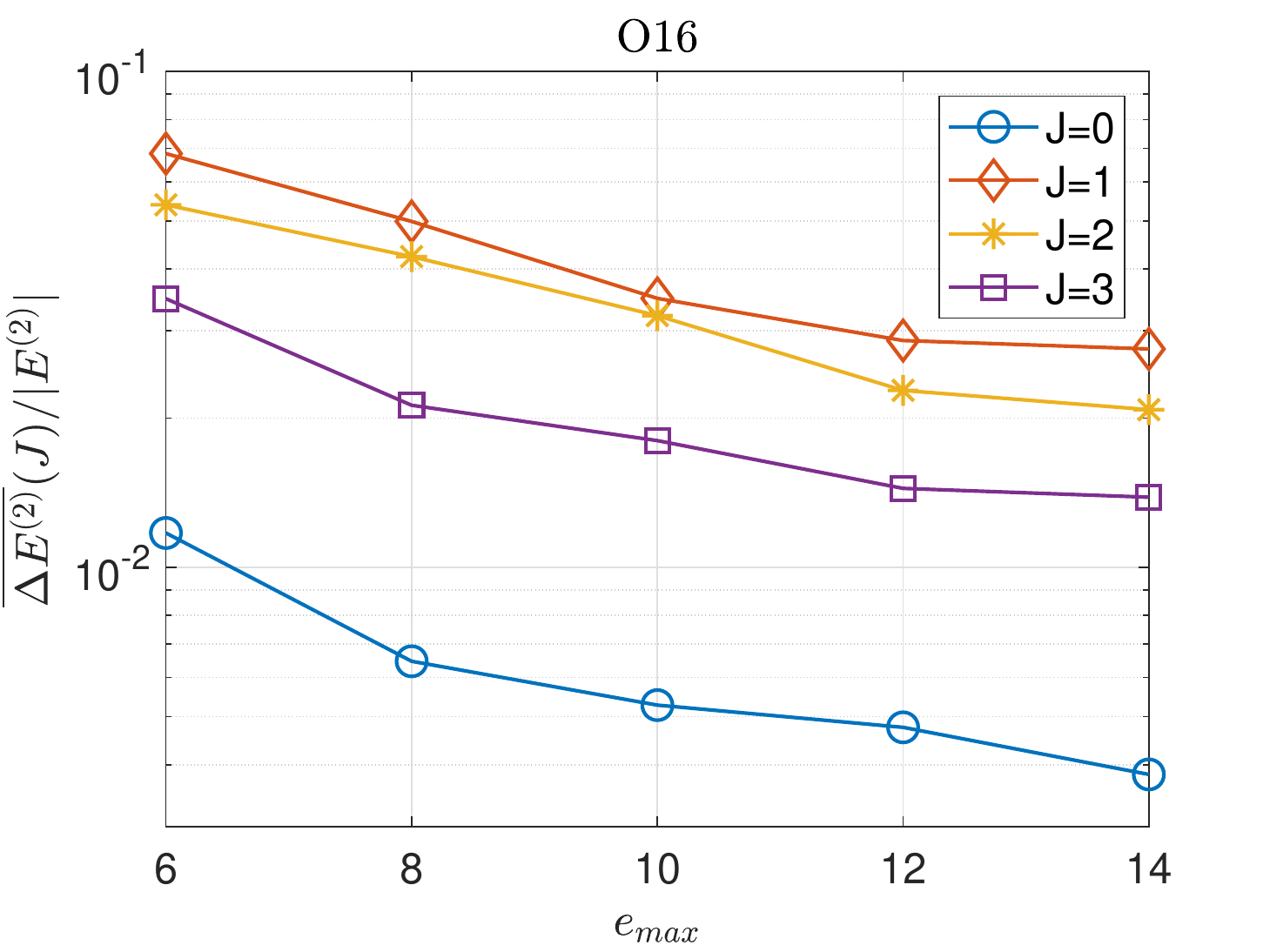}
    \end{subfigure}
    \begin{subfigure}{0.46\textwidth}
        \centering
        \includegraphics[width=0.9\linewidth]{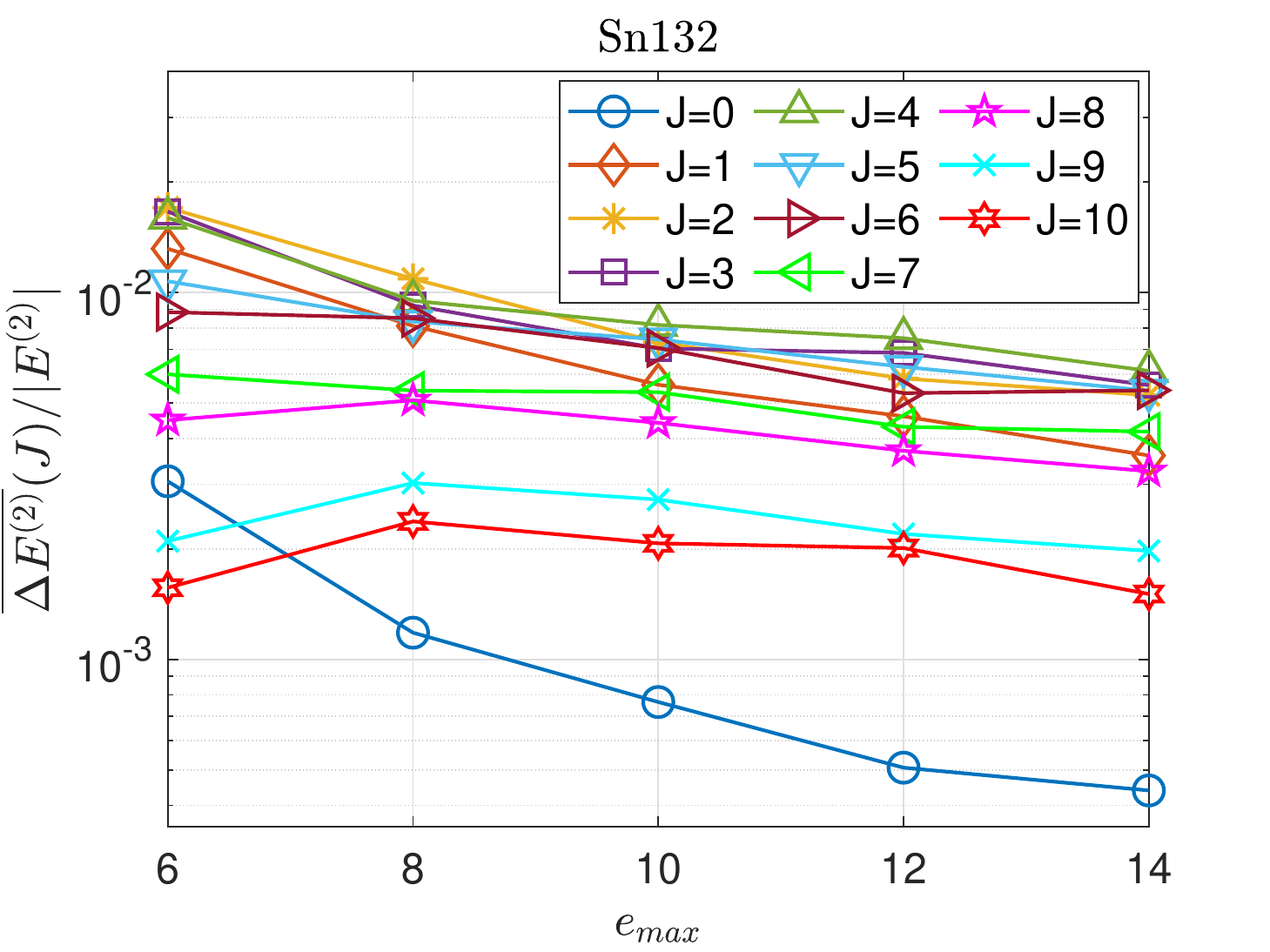}
    \end{subfigure}
    
    \caption{Breakdown of the mean relative errors $|\overline{\Delta E^{(2)}}/E^{(2)}|$ for $\ctot\leq 10^{-3}$ from Fig.~\ref{fig:deltaE_per-channel} by angular momentum channel.}
    \label{fig:deltaE_per-channel}
\end{figure*}

In Figure \ref{fig:error_vs_emax} we show the mean relative error of the second-order energy 
corrections as a function of the basis size parameter $\eMax$ for hundred- and thousandfold 
compressions of the Hamiltonian, considering $\nuc{O}{16}$ and $\nuc{Sn}{132}$ as typical 
examples. We see that for fixed compression, the error decays exponentially with the basis 
size in both nuclei, and this behavior is typical for all the nuclei and interactions we 
studied in this work --- results for additional nuclei are included in \ref{app:results}. 
There are weak fluctuations because of the random character of the JL embedding, and $\ctot$ 
is not strictly identical for each $\eMax$ and nucleus because of the varying dimensions (cf. 
Tab. \ref{tab:basis_size}). 

While some of the degrees of freedom of the large $\eMax$ basis 
are required to achieve converged results for the HF and MBPT ground-state energies, this 
result shows that an ever increasing amount of elements that are irrelevant for the
ground state are added to the Hamiltonian tensor as well. Consequently, we can use more
aggressive compressions if we are working in larger $\eMax$ spaces.

For fixed $\ctot$ and any given $\eMax$, the error for $\nuc{O}{16}$ is about twice as large 
as for $\nuc{Sn}{132}$. Since the fraction of relevant tensor elements is about an order of magnitude 
smaller for $\nuc{O}{16}$ than for $\nuc{Sn}{132}$ (cf. Fig. \ref{fig:num_pphh}), 
the random sampling performed by the JLE might be more likely to miss relevant contributions in
this nucleus. However, sparsity alone cannot be the main driver of this error, because we obtain
the expected exponential decay as the sparsity increases with $\eMax$. 

A breakdown of the mean relative errors by angular momentum is shown in Fig.~\ref{fig:deltaE_per-channel}. 
As we can see, the contributions $\overline{\Delta E^{(2)}(J)}$ all show the typical exponential
decay behavior, aside from the weak fluctuations discussed above. The relative impact of 
the channels is roughly correlated with the distribution of $N^{(2)}(J)$: Channels around $J=2$ or $3$ typically have
the highest number of relevant matrix elements and the largest contribution to the total $E^{(2)}$
in the nuclei we studied here. The $J=0$ channel is somewhat exceptional: It has a more restricted structure 
because the angular momentum 
selection rules $\eqref{eq:ang_mom_rule}$ force the single-particle angular momenta to be pairwise 
identical. Consequently, $N^{(2)}(0)$ is small, although channels with large $J$ have 
even smaller $N^{(2)}(J)$. The contributions of the high-$J$ channels to the overall error, however,
are amplified because they are weighted with $2J+1$ in Eq.~\eqref{equ:E2_total}, hence the $J=0$ channel contribution is consistently 
the smallest.

For $\nuc{O}{16}$, only channels with $J\leq 3$ contribute to the 
energy correction, while channels up to $J=10$ are relevant in $\nuc{Sn}{132}$ because orbitals 
with large single-particle angular momenta are occupied. The $\eMax=6$ results for $\nuc{Sn}{132}$ 
exhibit deviations from the exponential behavior in the large-$J$ channels that are
artifacts of the basis truncation, but their overall impact on the error is limited. 
The error contributions from the individual channels are consistently larger in $\nuc{O}{16}$
than in $\nuc{Sn}{132}$. Their distribution has a similar shape in both cases, although it 
is spread out more widely in the heavier nucleus. Thus, the greater overall error for $\nuc{O}{16}$
cannot be caused by strong discrepancies in the contributions from the low-$J$ and high-$J$ tails. 

\begin{figure}[t]
    \centering
    \includegraphics[width=0.9\linewidth]{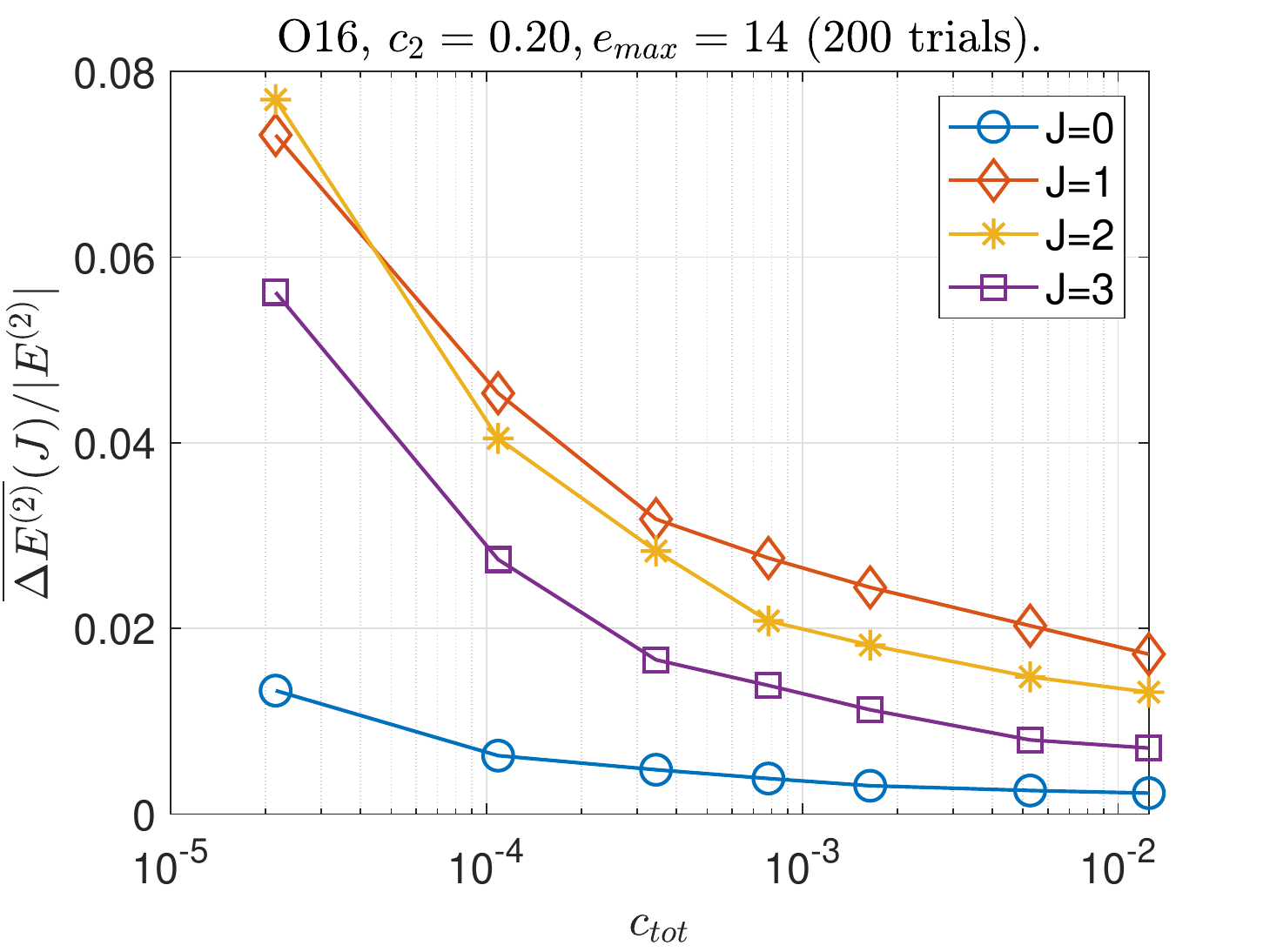}
    \caption{Contributions of each angular momentum channel to the mean relative error $\overline{\Delta E^{(2)}}/|E^{(2)}|$ in $\nuc{O}{16}$ as a function of the compression $\ctot$ (varying $c_1$  but keeping a fixed 
    compression $c_2=0.2$ for the second stage). Calculations were performed with the EM$1.8/2.0$ interaction in an $\eMax=14$ basis, and a two-stage JL embedding (RFD \text{+} RFD)\textsubscript{real} has been used.}
     \label{fig:deltaE_per-channel_vs-comp}
\end{figure}

In Fig.~\ref{fig:deltaE_per-channel_vs-comp}, we explore the behavior of $\overline{\Delta E^{(2)}(J)}/|E^{(2)}|$
for $\nuc{O}{16}$ when we keep $\eMax=14$ fixed and vary $\ctot$ instead. We find a smooth exponential 
growth of the error as we decrease $\ctot$ and make the compression more aggressive. For heavier
nuclei, we have more $J$ channels to consider but the behavior is very similar --- additional examples
are shown in the appendix.

Figure \ref{fig:err_vs_atom_emax8,14} summarizes the results from applying the two-stage 
(RFD \text{+} RFD)\textsubscript{real} JL embedding with $\ctot\leq 10^{-3}$ to the second-order 
energy corrections of several closed-shell nuclei. We see that we obtain the largest and smallest mean
relative errors for $\nuc{O}{16}$ and $\nuc{Sn}{132}$, respectively, while the other nuclei lie 
inbetween. The errors for $\nuc{Ca}{40,48}$ and $\nuc{Ni}{56}$ are rather similar at about
2\% for $\eMax=14$ bases that are tyically used in production-level calculations, and the errors for $\nuc{Ni}{78}$ and $\nuc{Sn}{132}$ in the
same basis size are in the 1-2\% range.

Overall, we see that the mean relative error decreases with the particle number $A$. The large
jump between $\nuc{O}{16}$ and the calcium isotopes could indicate a ``shell effect'' as
the occupation of orbitals with growing single-particle $j$ also implies that higher $J$ channels
of the Hamiltonian can contribute. Applying the method to some $sd$-shell nuclei in the future
could help clarify how smooth the $A$ (or the $N$ and $Z$ dependencies) are, but the absence
of candidate nuclei with strong shell closures means that we will have to switch to a more general 
form of MBPT. 

\begin{figure}[t]
    \centering
    \includegraphics[width=0.45\textwidth]{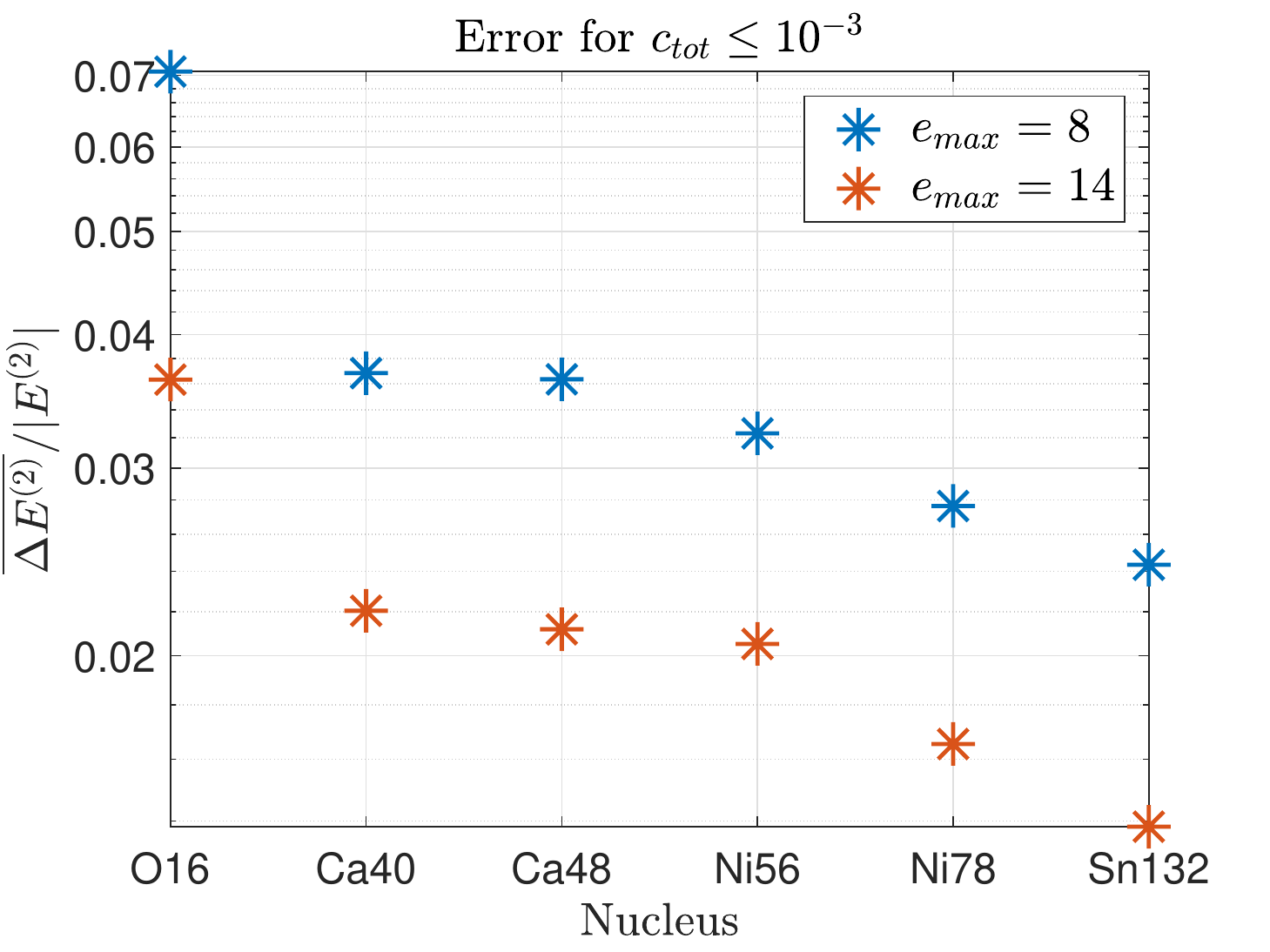}
    \caption{Mean relative error $\overline{\Delta E^{(2)}}/|E^{(2)}|$ for several closed-shell nuclei,
    using $\eMax=8,14$ bases and a compression $c_{tot} \leq 10^{-3}$. All results were obtained 
    with the EM$1.8/2.0$ interaction, using the (RFD \text{+} RFD)\textsubscript{real} embedding
    and carrying out $200$ trials.}
    \label{fig:err_vs_atom_emax8,14}
\end{figure}

\subsubsection{Interaction Dependence}

The next aspect we want to explore is the performance of the JLE for interactions with different
resolution scales. For this purpose, we have applied the (RFD \text{+} RFD)\textsubscript{real}
embedding in calculations with other members of the EM$\lambda/\Lambda$ family of interactions
\cite{Hebeler:2011dq}, specifically EM2.8/2.0 and EM2.0/2.5.

By varying the resolution scale $\lambda$ through SRG evolution (and readjusting 
the cutoff $\Lambda$ and low-energy constants of the 3N interaction), correlations are re-shuffled 
between the Hartree-Fock reference state and 
the perturbative corrections to the wave function. As we can see for the example of $\nuc{O}{16}$ 
in Table \ref{tab:O16_energies}, the Hartree-Fock energy $\Eref$ and the second-order correction
$E^{(2)}$ vary by factors 3--4 for the three interactions we consider here. Note, however, that
the total MBPT(2) energy only changes by about 10\%. For EM1.8/2.0, the bulk of the ground-state 
energy is already obtained at the mean-field level, but $E^{(2)}$ is still a sizeable correction 
of greater than 30\%, so third order corrections are usually checked to establish convergence of 
the series expansion (or lack thereof). The same kind of checks are absolutely mandatory for 
EM2.8/2.0, since it yields an $E^{(2)}$ that is more than three times greater than $\Eref$.

\begin{table}[b]
    \centering
    \begin{tabular}{|c|c|c|c|}
    \hline\hline
    Interaction & $\Eref\;[\MeV]$ & $E^{(2)}\,[\MeV]$ & $E\;[\MeV]$ \\ 
    \hline
    EM1.8/2.0 &  -90.29 & -33.43 & -123.72 \\
    EM2.0/2.5 &  -68.78 & -44.92 & -113.70 \\
    EM2.8/2.0 &  -26.20 & -83.23 & -109.43 \\
    \hline
    \end{tabular}
    \caption{
        Hartree-Fock ($\Eref$), second-order MBPT correction ($E^{(2)}$) and total MBPT(2) energy $E=\Eref + E^{(2)}$
        of $\nuc{O}{16}$ for three interactions from the EM$\lambda/\Lambda$ family. All calculations were performed
        with $\eMax=14$ at optimal $\hbar\omega$.
    }
    \label{tab:O16_energies}
\end{table}

\begin{figure}[t]
    \centering
    \includegraphics[width=0.45\textwidth]{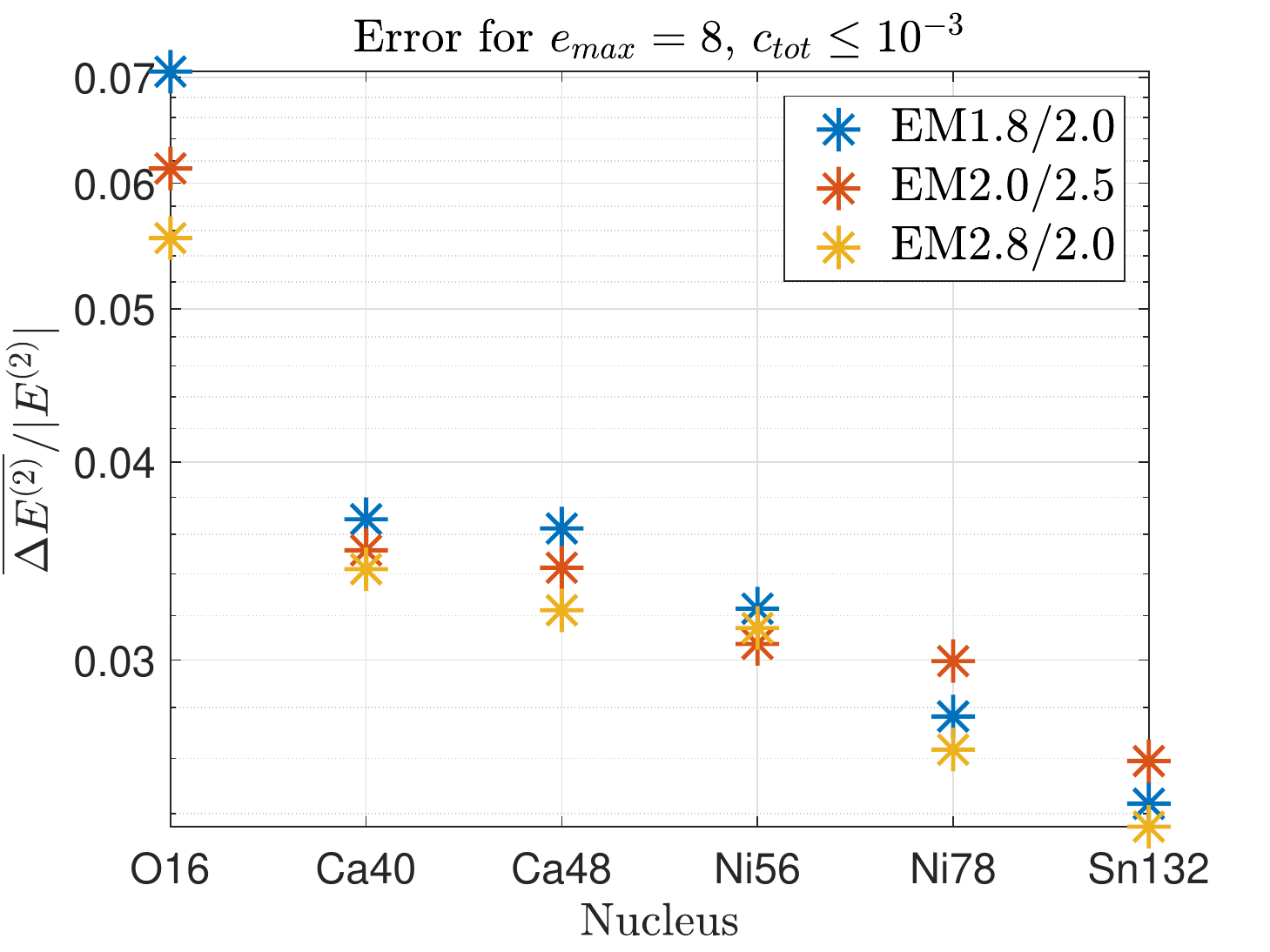}
    \caption{Mean relative error $\overline{\Delta E^{(2)}}/|E^{(2)}|$ in various closed-shell
    nuclei for different members of the EM$\lambda/\Lambda$ interaction family. All
    calculations were performed with an $\eMax=8$ basis and compression $\ctot \leq 10^{-3}$, 
    using the (RFD \text{+} RFD)\textsubscript{real} embedding and $200$ trials.}
    \label{fig:err_vs_atom_emax8}
\end{figure}

In Figure \ref{fig:err_vs_atom_emax8}, we show $\overline{\Delta E^{(2)}}/|E^{(2)}|$ 
that result from appying the (RFD \text{+} RFD)\textsubscript{real} embedding to evaluate 
$E^{(2)}$ for our three interactions. Despite the significant differences in resolution 
scales and the resulting size differences in $E^{(2)}$, the mean relative errors are very similar.
Since calculations with interactions like EM2.8/2.0 typically force us to use large single-particle 
bases to reach convergence, it is a very welcome result that we will be able to use JLEs
for compression with reliable uncertainties in such applications.

\subsubsection{Total Energy}

\begin{figure}[t]
    \centering
    \includegraphics[width=0.45\textwidth]{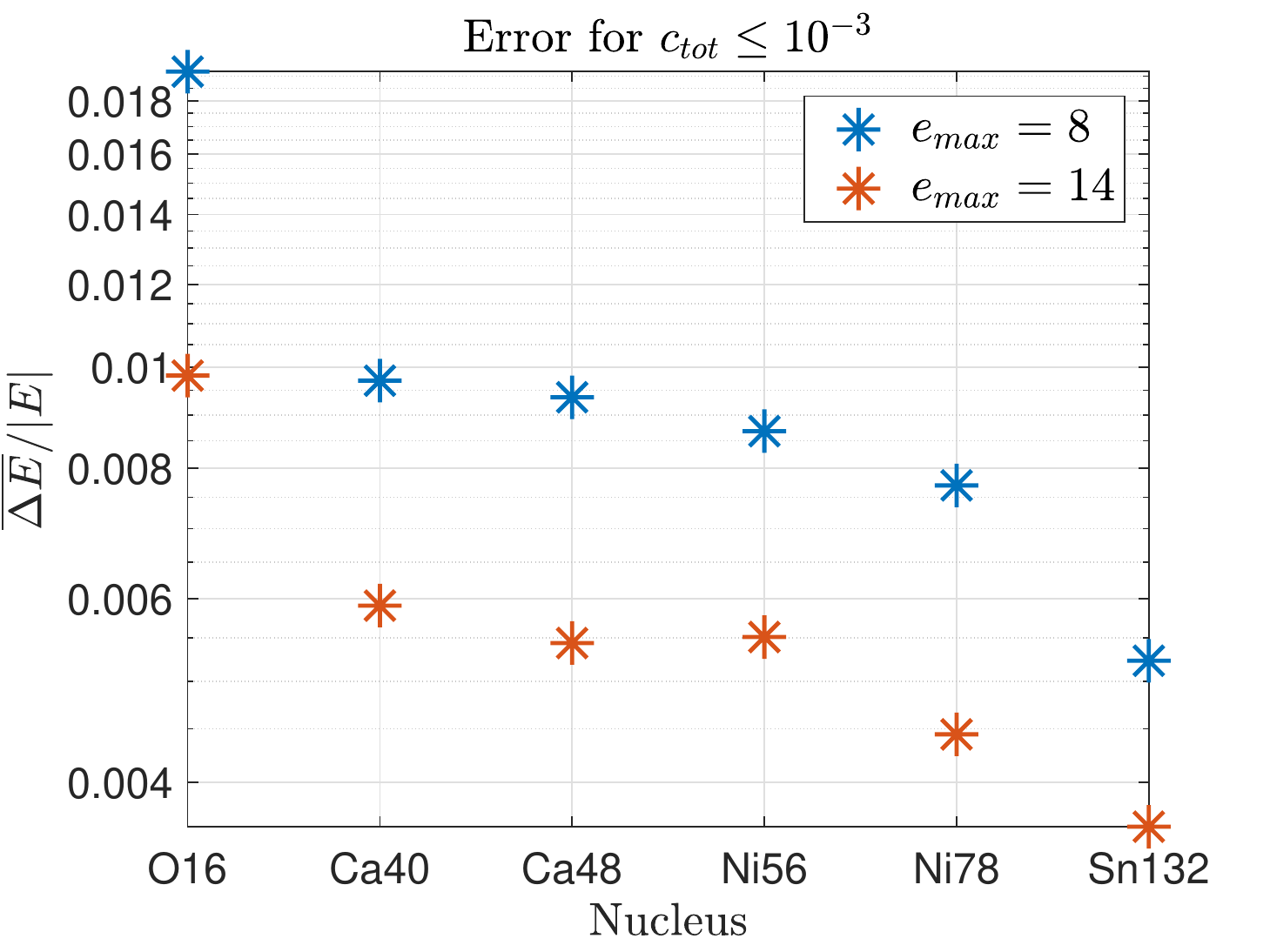}
    \caption{
        Mean relative error of the total energy, $\overline{\Delta E}$, for closed-shell nuclei, using
        single-particle bases with $\eMax=8,14$ (at optimal $\hbar\omega$) and compression $c_{tot} \leq 10^{-3}$. All calculations were performed with the EM$1.8/2.0$, using (RFD \text{+} RFD)\textsubscript{real} and $200$ trials.
    }
    \label{fig:err_vs_atom_emax8,14_otherHW}
\end{figure}

While the previous sections have presented a detailed analysis of JLEs in the computation of 
$E^{(2)}$, the quantity that is ultimately relevant for comparison with experimental data is 
the total energy $E=\Eref + E^{(2)}$ (cf. Sec.~\ref{sec:mbpt} and \ref{sec:measures}). In Fig. \ref{fig:err_vs_atom_emax8,14_otherHW}, 
we show the mean relative errors of $E$ that result from applying the (RFD \text{+} RFD)\textsubscript{real}
embedding with $\ctot\leq 10^{-3}$. Since $\Eref$ is determined prior to application of the JLE, 
this merely amounts to
a propagation of the mean absolute errors in $E^{(2)}$ to the total energy. Consequently,
the behavior of the errors is very similar to what we found in Fig.~\ref{fig:error_vs_emax}, but
their size is reduced: The largest errors are still incurred in $\nuc{O}{16}$, with 2\% for $\eMax=8$
and 1\% for $\eMax=14$. For all other nuclei, even in the small basis the errors are well below
1\%, and therefore much smaller than current systematic uncertainties due to the truncation of the
perturbation series, NO2B approximation, or the parameters of the input interactions (see, e.g., Ref. \cite{Hergert:2020am}).
Thus, JLEs can be used to greatly accelerate large-scale exploratory calculations that seek to 
quantify these uncertainties.

\begin{figure}[t]
    \centering
    \includegraphics[width=0.45\textwidth]{{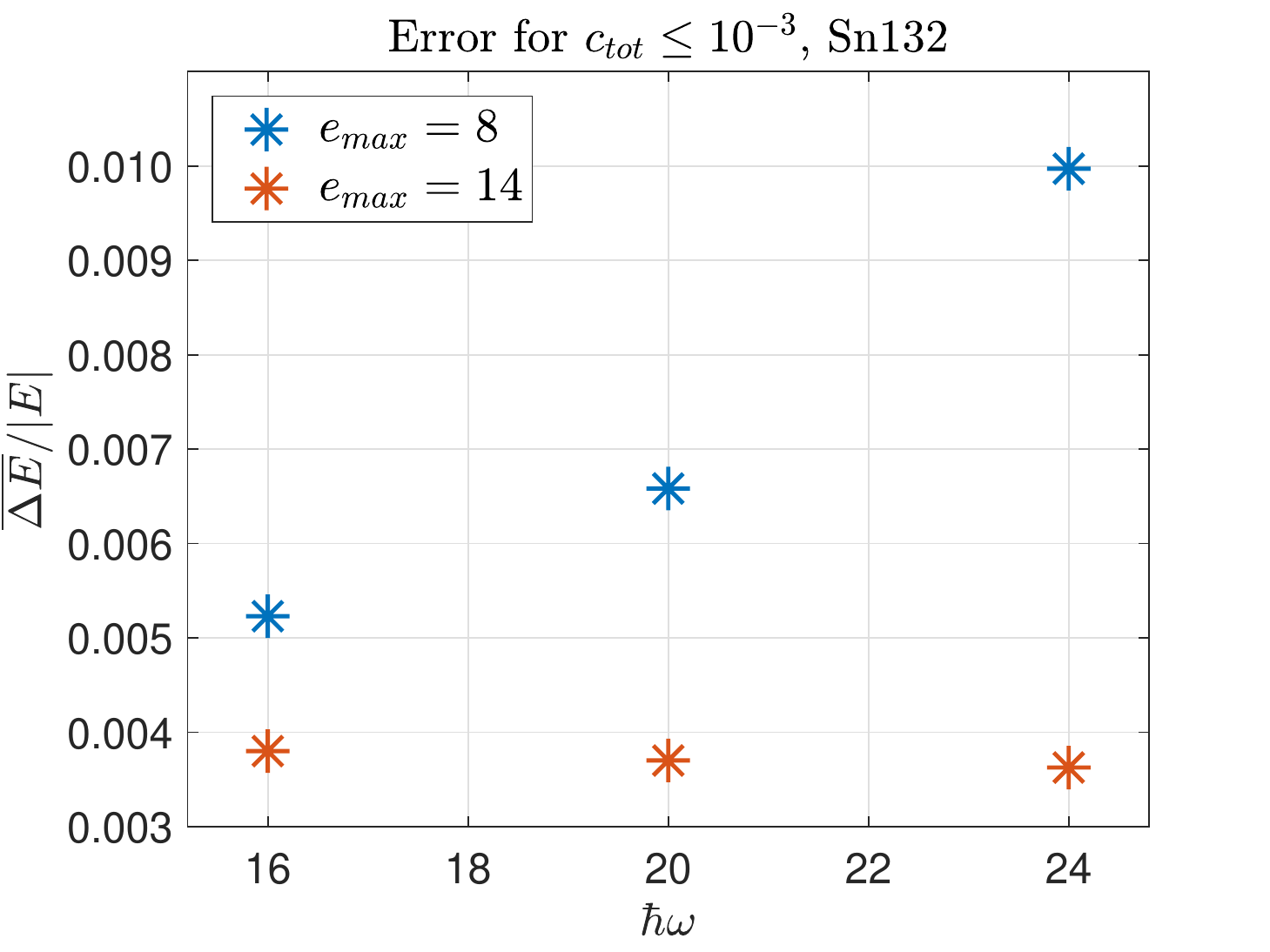}}
    \caption{Mean relative error $\overline{\Delta E}/|E|$ of $\nuc{Sn}{132}$ as a function of $\hbar\omega$ for basis 
    sizes $\eMax=8,14$ and compression $c_{tot} \leq 10^{-3}$. All results are obtained with EM$1.8/2.0$, using  
    (RFD \text{+} RFD)\textsubscript{real} and $200$ trials.}
    \label{fig:err_vs_hw_emax8_Sn132}
\end{figure}

Since the starting point for our Hartree-Fock and subsequent MBPT(2) calculations are interaction
matrix elements in a spherical harmonic oscillator basis (cf. Sec.~\ref{sec:bases}), our results
will in general retain some dependence on the oscillator parameter $\hbar\omega$ because of the
basis' finite size. In Fig.~\ref{fig:err_vs_hw_emax8_Sn132}, we explore this dependence for $\nuc{Sn}{132}$, 
since this nucleus requires the largest basis to achieve reasonable converge. The energy correction
$E^{(2)}$ exhibits only a weak $\hbar\omega$ dependence: For instance, with $\eMax=8$ and using the 
EM1.8/2.0 interaction, it varies from $-220\,\MeV$ at $\hbar\omega=16\,\MeV$ to $-210\,\MeV$ at 
$\hbar\omega=24\,\MeV$. Meanwhile, the mean-field energy $\Eref$ varies by $450\,\MeV$ in this window. 
This variation causes the typical parabolic shape that we also observe ground-state energy
convergence plots for Hartree-Fock and other many-body approaches, and it leads to
significant changes in the mean relative error $\Delta E/|E|$ as well, which ranges from 0.5\%
to 1\%. As we increase the basis size to $\eMax=14$, $\Eref$ is much better converged, and the
error settles in at about 0.36\%, consistent with Fig.~\ref{fig:err_vs_atom_emax8,14}.

Exploring the sensitivity of energies and other observables to variations of $\eMax$ and $\hbar\omega$ 
is standard practice for assessing the convergence of nuclear many-body calculations, and our present 
findings indicate that JLEs can be integrated into such analyses in a straighforward fashion.

\subsection{Radii}
\label{sec:radius_corrections}
While ground-state energies are typically the main quantity of interest in MBPT calculations, we
can use the perturbative corrections to the wave function to evaluate other observables like the 
mean-square radius, as explained in Section \ref{sec:mbpt}. Formally, the leading correction to 
the mean-square radius is a first-order contribution from the two-body part of $R$ (cf. Eq.~\eqref{eq:def_R}). 
Upon evaluation, we find that its size is on the order of 0.01\% of the reference state expectation 
value of the operator: For instance, for $\nuc{Ca}{40}$ with $\eMax=14$, $\hbar\omega=16$, and the EM1.8/2.0 
interaction, the mean-field mean-square radius is $R_0=9.98\,\fm^2$, and $R^{(1)}_2 = 0.003\,,\fm^2$. 
The second-order corrections from the one-body operators, on the other hand, are $R^{(2)}_1 = 1.40\,\fm^2$, 
and generally on the order of 10-15\% for the nuclei studied here. For this reason, we do not consider 
$R^{(1)}_2$ and only focus on the one-body contributions in the following discussion.

\begin{figure}[t]
    \centering
    \includegraphics[width=0.45\textwidth]{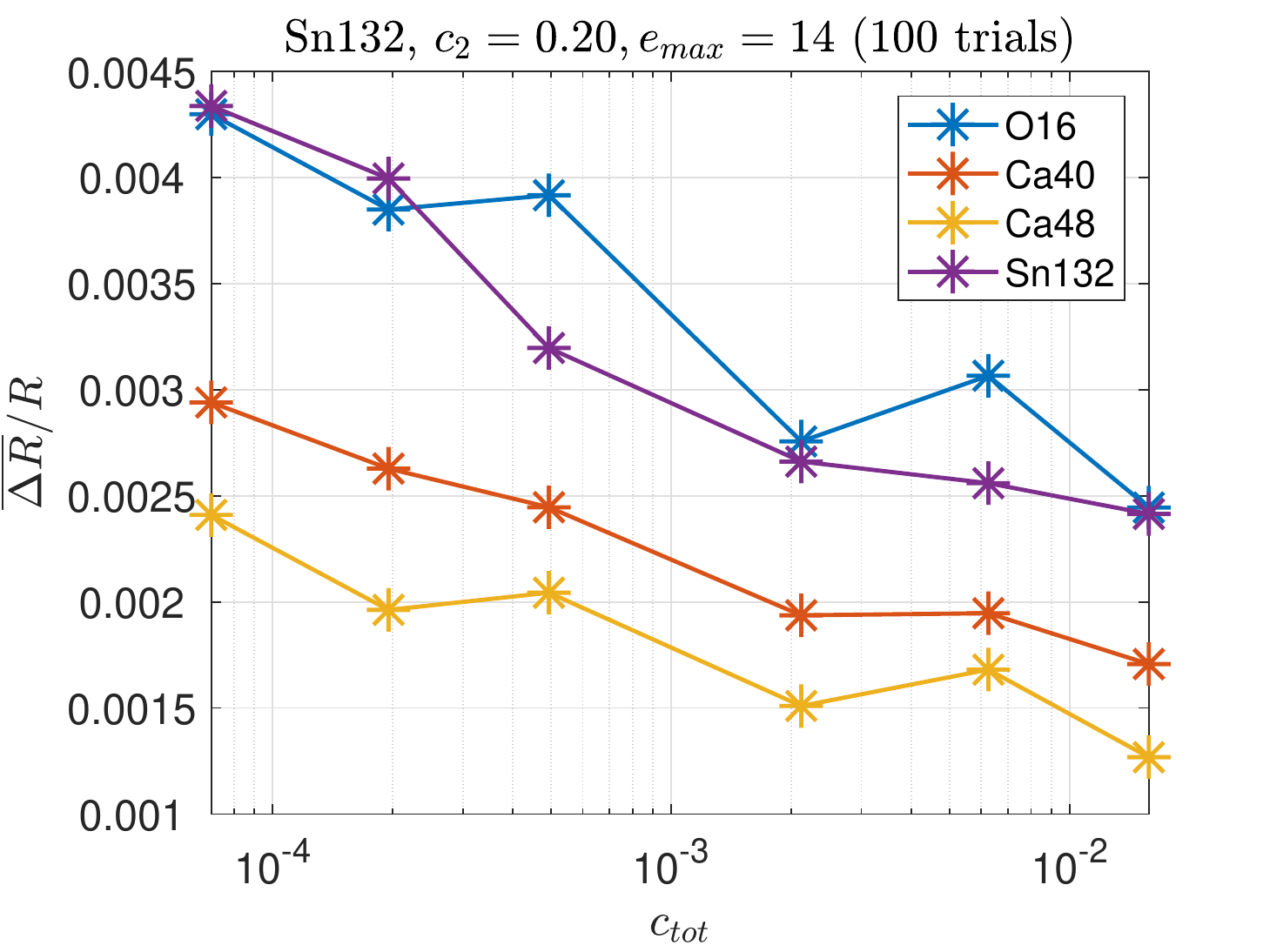}
    \caption{Mean relative error of the total radius correction as a function of the compression 
    (relative to the total radius correction) for various closed-shell nuclei, using
    the EM$1.8/2.0$ interaction and $\eMax=14$. All results were obtained with the 
    (RFD \text{+} RFD)\textsubscript{real} embedding and 100 trials, varying $c_1$ but
    keeping $c_2=0.2$ fixed for the second RFD stage.}
    \label{fig:Radius-tot-err_vs_comp_emax14}
\end{figure}

Implementing the (RFD \text{+} RFD)\textsubscript{real} embedding for $R^{(2)}_1$ according to Sec.~\ref{sec:JLapplication},
we obtain the mean relative errors $\overline{\Delta R}/|R|$ shown in Fig.~\ref{fig:Radius-tot-err_vs_comp_emax14}.
For all the closed-shell nuclei considered here, the errors decay exponentially (up to fluctuations
due to the random sampling), just like in the case of the energies. For target compressions of $\ctot\leq 10^{-3}$
to $10^{-2}$ that we discussed before, the errors are on the order of 0.2-0.3\%, i.e., the JLEs work
even better for the radii than for the energies. The error is once again largest in $\nuc{O}{16}$,
which is expected based on the discussion in Sec.~\ref{sec:energy_corrections}. For $\nuc{Ca}{40,48}$,
the values shrink significantly, but the error for $\nuc{Sn}{132}$ is comparable to that of $\nuc{O}{16}$. 
The most likely explanation is that the situation is analogous to what we found in Fig.~\ref{fig:err_vs_hw_emax8_Sn132},
and that the radius of $\nuc{Sn}{132}$ is not sufficiently well converged --- 
radii generally have a slower convergence in the basis size $\eMax$ than energies (see, e.g., \cite{Hergert:2016jk}).
Even so, Fig.~\ref{fig:Radius-tot-err_vs_comp_emax14} shows that the (mean) relative errors
caused by applying JLEs to the evaluation of nuclear radii are negligible compared to other
sources of error.

\section{Conclusions}\label{sec:conclude}
In the present work, we have initiated a program to explore the use of modewise 
Johnson-Lindenstrauss embeddings (JLEs) as a compression tool for nuclear many-body 
theory. Applying such embeddings to the calculation of ground-state energies in second-order
Many-Body Perturbation Theory (MBPT(2)), we were able to compress the Hamiltonian in
a large-basis calculation more than thousandfold while only incurring errors below
1\%, and we found that the mean relative errors caused by the JLE behave very regularly
across single-particle basis sizes and the angular momentum channels of our $J$-scheme
calculations. 

The memory savings we achieved through JLE-based compression are comparable to those
of an a priori, theory-based selection of Hamiltonian tensor entries that can contribute
to the second-order energy and first-order wave function. This means that despite its
randomized and therefore data-oblivious nature, the JLE captures the relevant physics with very high accuracy without 
prior assumptions about the structure of the Hamiltonian or the occupancy of the orbitals 
in all the nuclei we studied here. We also stress that the compressed Hamiltonian can be 
readily reused in future many-body applications, although the embedding errors will have 
to be reassessed if it serves as input for methods other than MBPT(2).

An obvious next step is to apply the JLEs in third and higher orders of MBPT, where the
compression of the Hamiltonian will also enable order-of-magnitude reductions in computing 
time as an additional benefit. Since 
higher-order energy and wave-function corrections probe all elements of the Hamiltonian,
we anticipate somewhat larger errors for any given compression than in MBPT(2). However,
the contributions of these corrections to the ground-state energy and other observables 
get progressively smaller if the many-body perturbation series converges, which will 
counteract the growing compression error. We already saw examples of this behavior in the
present work: The mean relative error of the total ground-state energy $E$ is smaller than
that of the second order correction $E^{(2)}$, which is an $O(10\%)$ correction, and the
third order correction is typically of order $O(1\%)$. Similarly, the leading correction
for the two-body radius operator has a much greater compression error than the energy, but 
it is a negligible correction to the total radius expectation value and therefore the
error does not matter.

Eventually, we intend to apply the JLE in nonperturbative approaches like the In-Medium
Similarity Renormalization Group or the Coupled Cluster method. Their working equations 
consists of expressions with similar 
complexity as third (and possibly higher) order MBPT, but they must be evaluated 
iteratively. Thus, the computing time savings due to the use of JLEs will be amplified
in such applications.

In parallel to pursuing applications of JLEs in more sophisticated many-body approaches,
we will deploy them in $M$-scheme calculations with symmetry unrestricted single-particle 
bases that are relevant for the description of nuclei with complex intrinsic structures. 
For such bases, the dimension for each tensor index is at least an order of magnitude
larger than in the $J$-scheme case discussed in the present work (cf. Sec.~\ref{sec:bases}). 
Naively, this should allow us to apply more agressive compression schemes based on our
observations for growing $\eMax$ in Sec.~\ref{sec:JLapplication}. The relevant tensors 
will also be sparser due to additional selection rules being in effect, hence we must 
assess the performance of the random sampling performed by the JLEs under these conditions. Moreover,
the $M$-scheme calculations will typically yield many-body wave functions with broken
symmetries that need to be restored explicitly to make accurate comparisons with experimental
data. Since symmetry restoration techniques rely on a delicate balance of interaction 
and wave function contributions, we will have to carefully study their interplay with 
the JLEs. To properly deal with all these aspects of future $M$-scheme applications 
we might have to design JLE schemes that incorporate features of the underlying physics 
more explicitly. 

Last but not least, we will explore the use of JLEs in the recently launched efforts
to factorize nuclear interaction tensors and many-body methods in order to control
their computational scaling. The essential numerical steps are typically cast in the
form of large-scale optimization problems that are solved with alternating
least-squares methods \cite{Tichai:2022rq,Tichai:2019xz,Schutski:2017xr,Parrish:2019zg}, 
and dimensional reduction via modewise JLEs enables 
order-of-magnitude speedups in such calculations \cite{doi:10.1137/17M1112303,sun2020sketching,Iwen:2021my}.

\section*{Data Availability}
The simulation codes and data for $\nuc{Sn}{132}$ obtained with the EM$1.8/2.0$ interaction in an 
$\eMax=6$ basis can be accessed in a Github repository at \url{https://github.com/azarerepo/JL4MBPT}. 
Data for other nuclei and basis sizes are available from the authors upon request.

\section*{Acknowledgments} 
C. A. Haselby and M. Iwen were supported in part by the National Science Foundation under Award No. DMS 2106472. H. Hergert and R. Wirth acknowledge support by the U.S. Department of Energy, Office of Science, Office of Nuclear Physics under Awards No. DE-SC0017887 and DE-SC0018083 (NUCLEI SciDAC-4 Collaboration.)

\appendix

\section{Additional Results}
\label{app:results}

For completeness, we compile detailed results for energy corrections in additional nuclei in this appendix. 

\begin{figure}[t]
    \centering
    \begin{subfigure}{0.46\textwidth}
        \centering
        \includegraphics[width=0.9\linewidth]{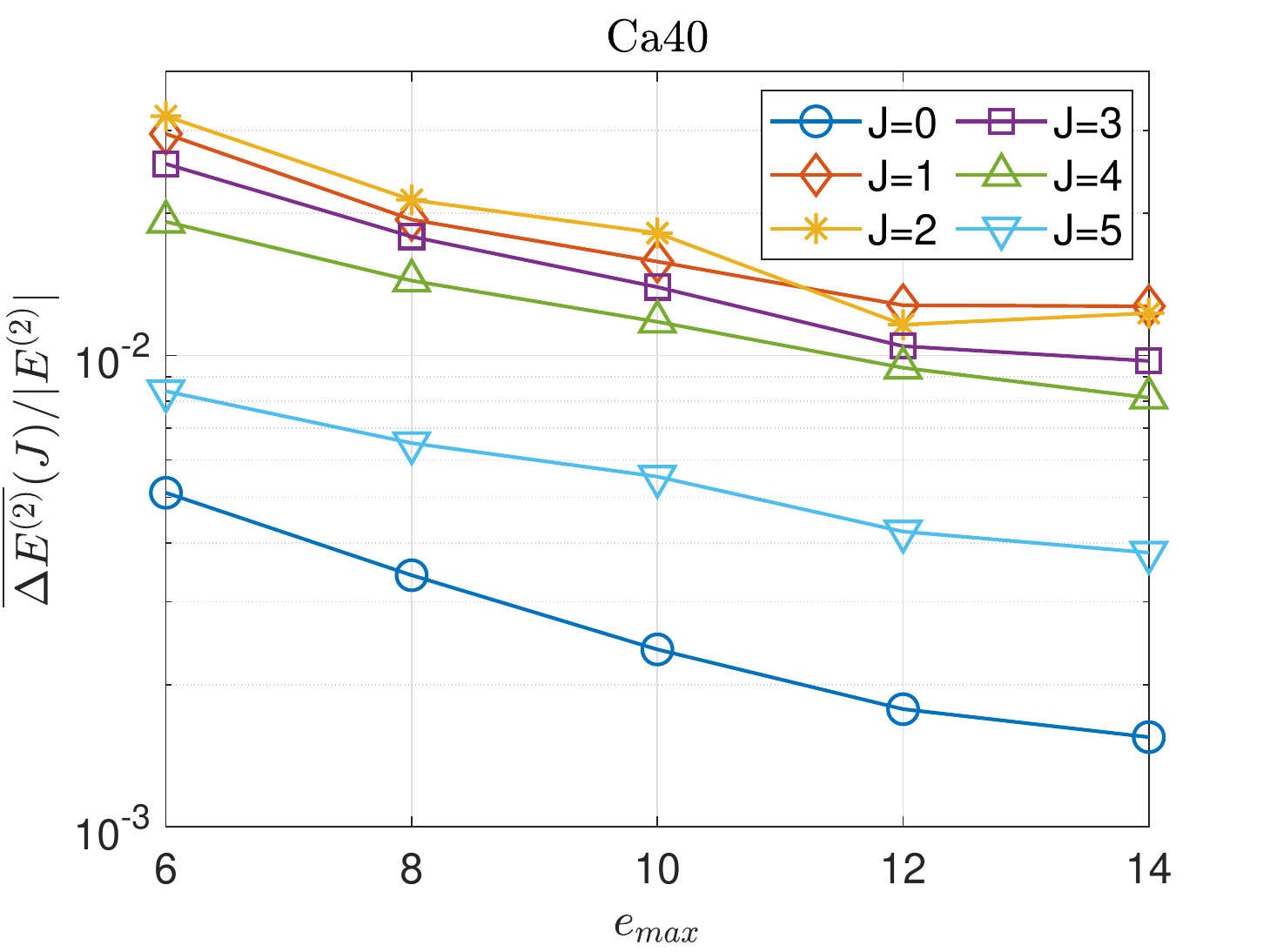}
    \end{subfigure}
    \begin{subfigure}{0.46\textwidth}
        \centering
        \includegraphics[width=0.9\linewidth]{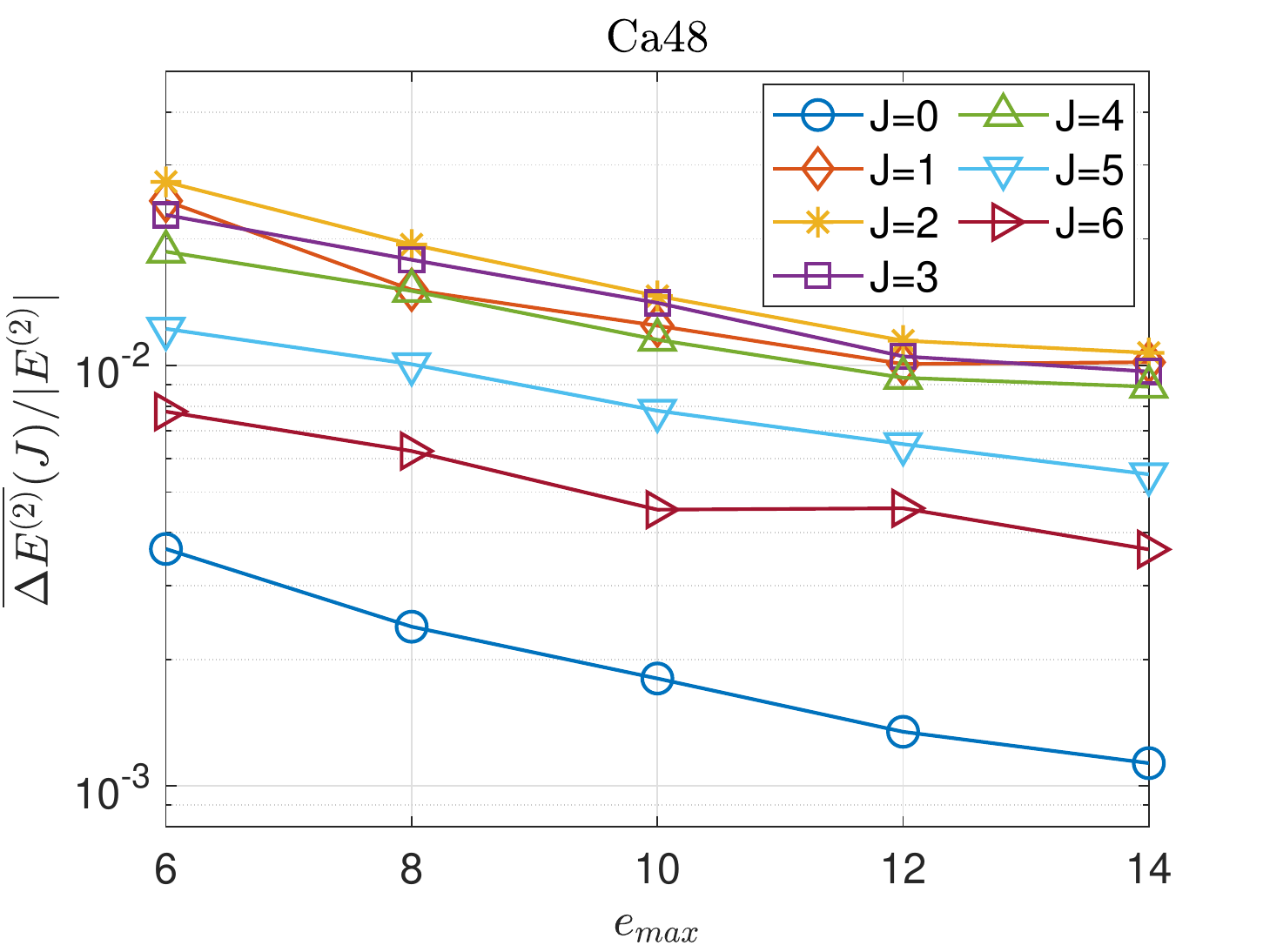}
    \end{subfigure}

    \caption{Contributions of each angular momentum channel to the mean relative error $\overline{\Delta E^{(2)}}/|E^{(2)}$ as a function of $\eMax$, for fixed compression $c_{tot} \leq 10^{-3}$. All calculations were performed with the EM$1.8/2.0$ interaction. Our favored two-stage JL embedding (RFD \text{+} RFD)\textsubscript{real} (cf.~Sec.~\ref{sec:JLscheme}) was used for compression, and $200$ trials were carried out. 
    }
    \label{fig:deltaE_per-channel_appendix}
    \medskip
\end{figure}

Figure \ref{fig:deltaE_per-channel_appendix} shows the angular-momentum channel breakdown of the mean 
relative error of $E^{(2)}$ for different basis sizes in the nuclei $\nuc{Ca}{40,48}$. The observed
trends match what we found and discussed for $\nuc{O}{16}$ and $\nuc{Sn}{132}$ via
Fig.~\ref{fig:deltaE_per-channel}.

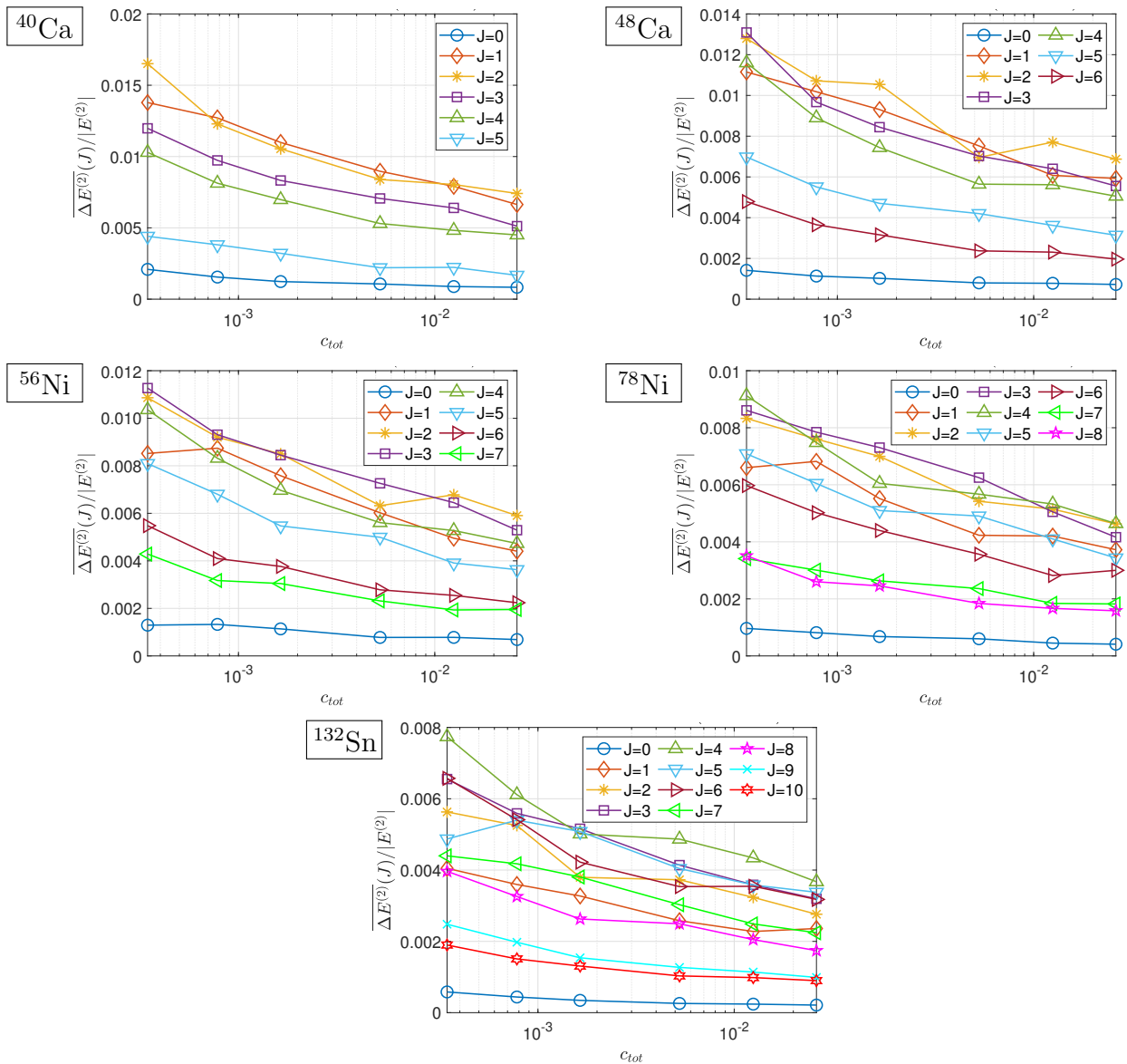
\begin{figure*}[t]
    \centering

    \setlength{\unitlength}{\textwidth}
    \begin{picture}(1.0000,0.89000)
        \put(0.0750,0.6000){\includegraphics[width=0.41\unitlength]{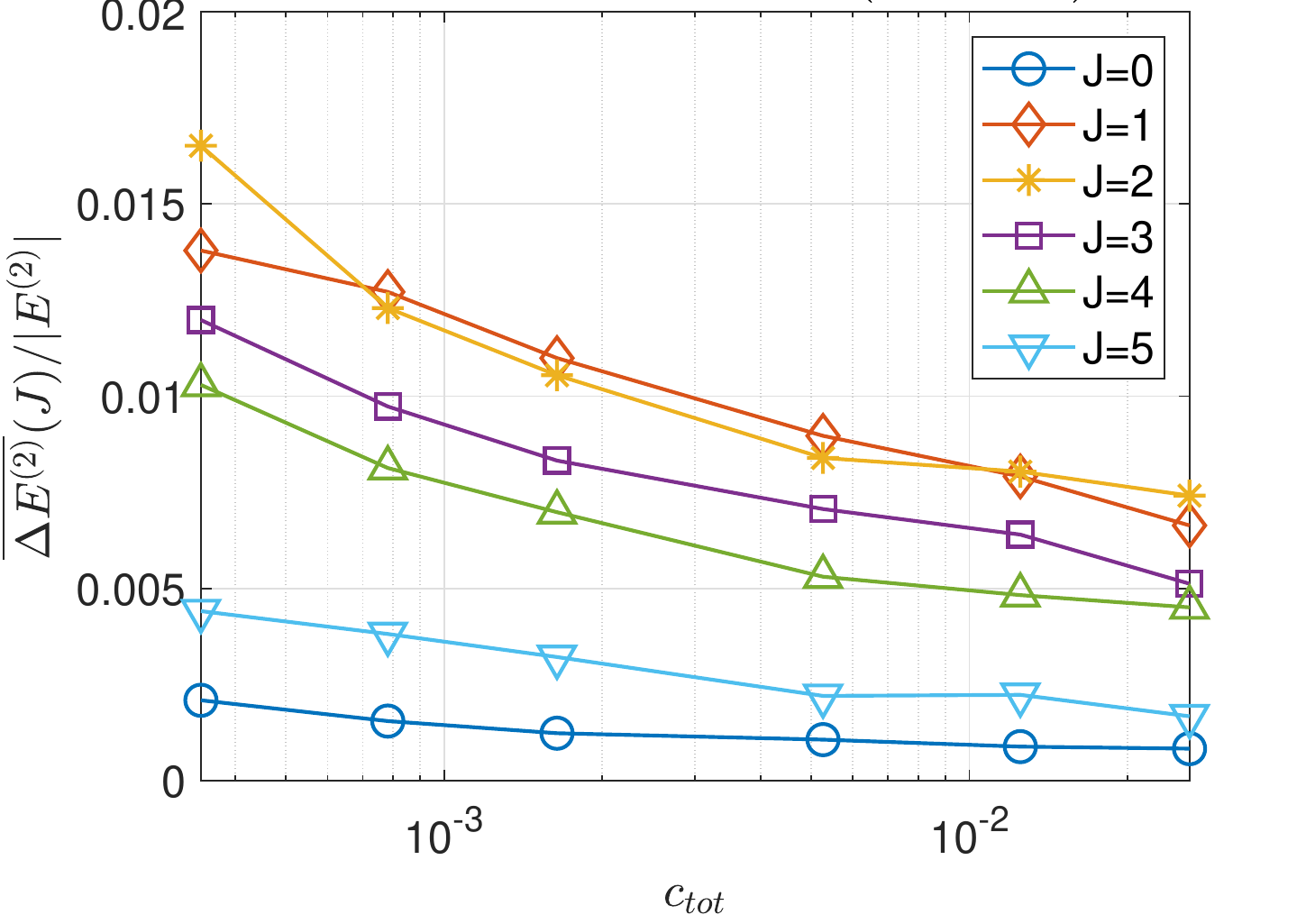}}
        \put(0.5750,0.6000){\includegraphics[width=0.41\unitlength]{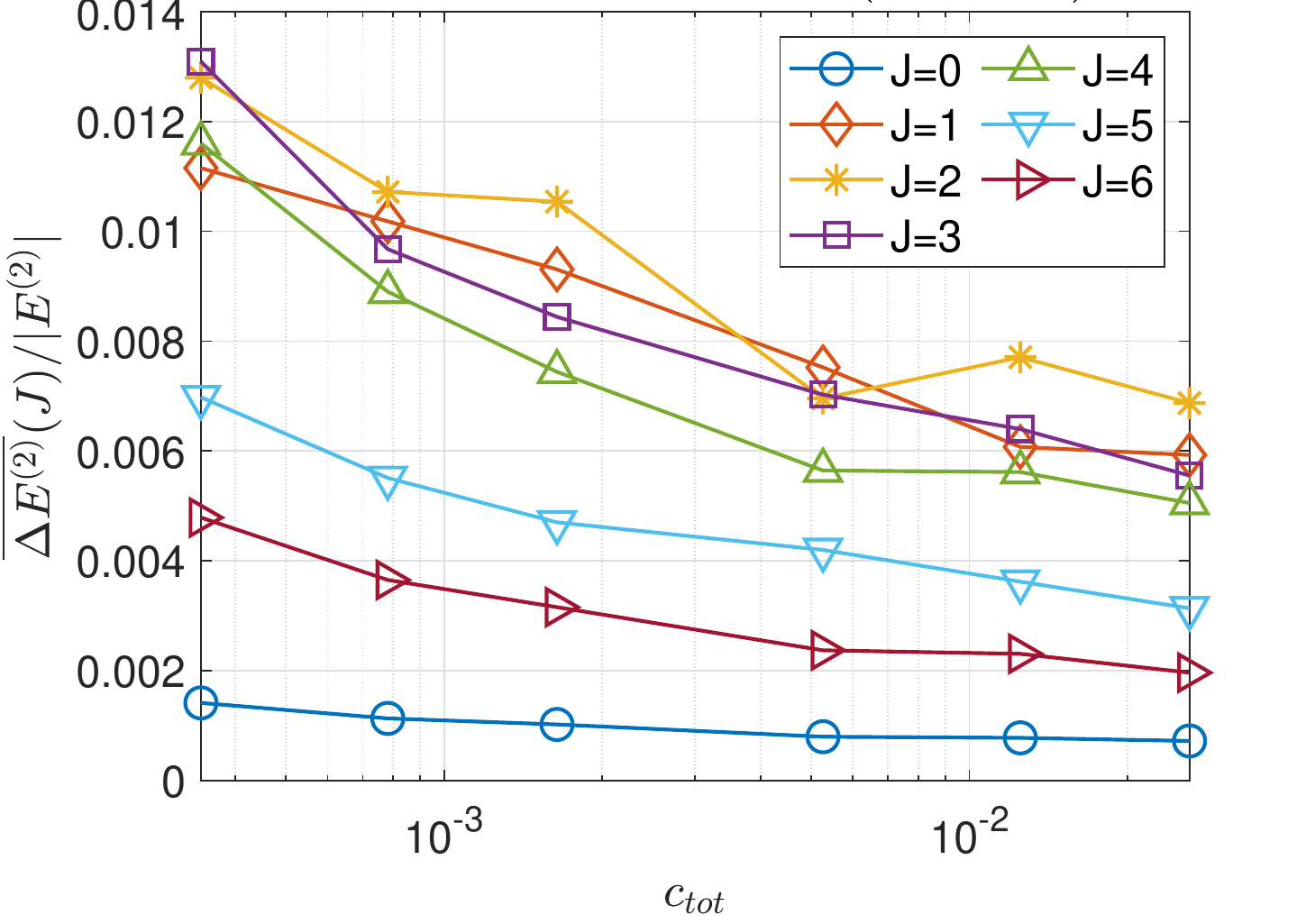}}
        \put(0.0750,0.3000){\includegraphics[width=0.41\unitlength]{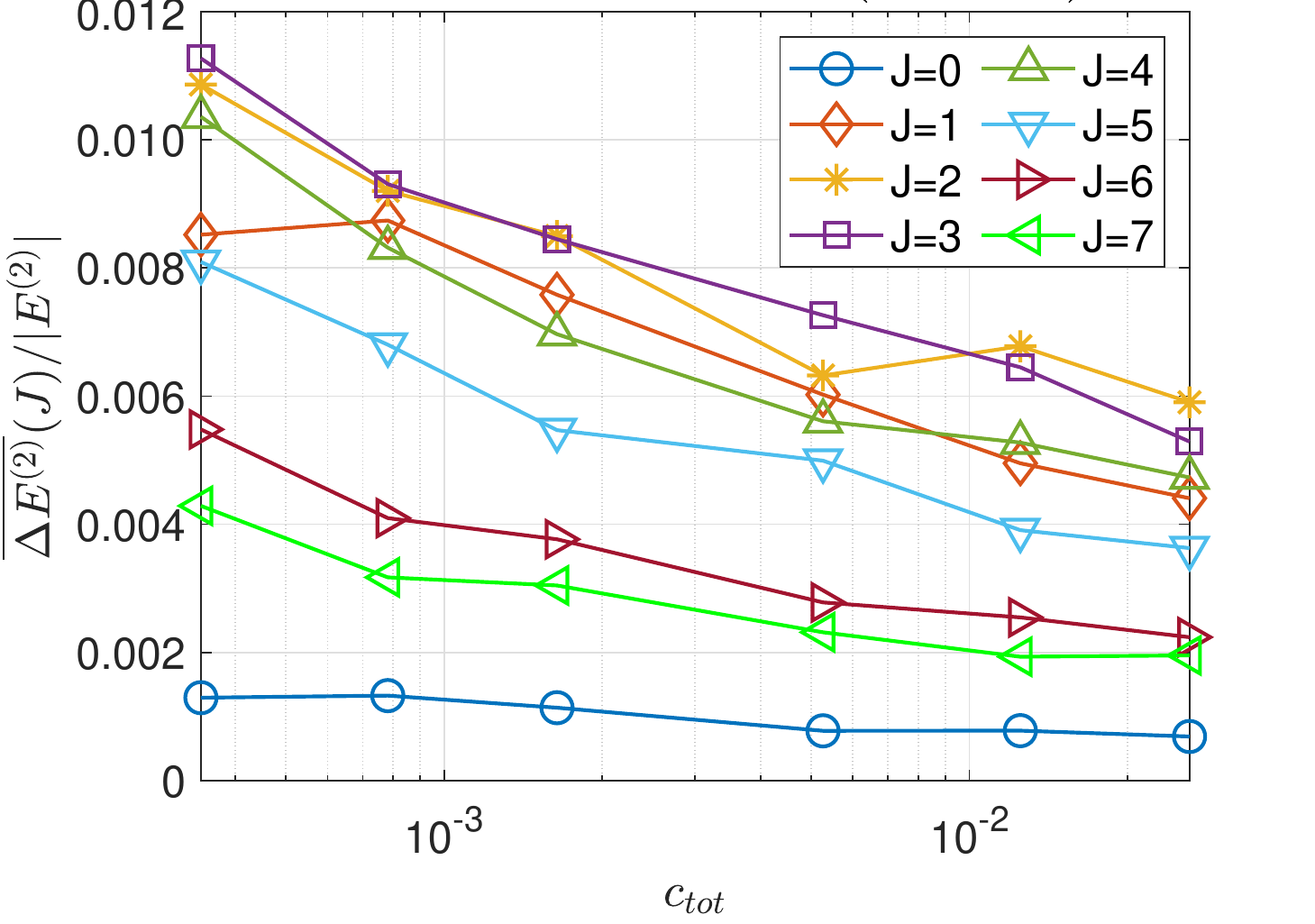}}
        \put(0.5750,0.3000){\includegraphics[width=0.41\unitlength]{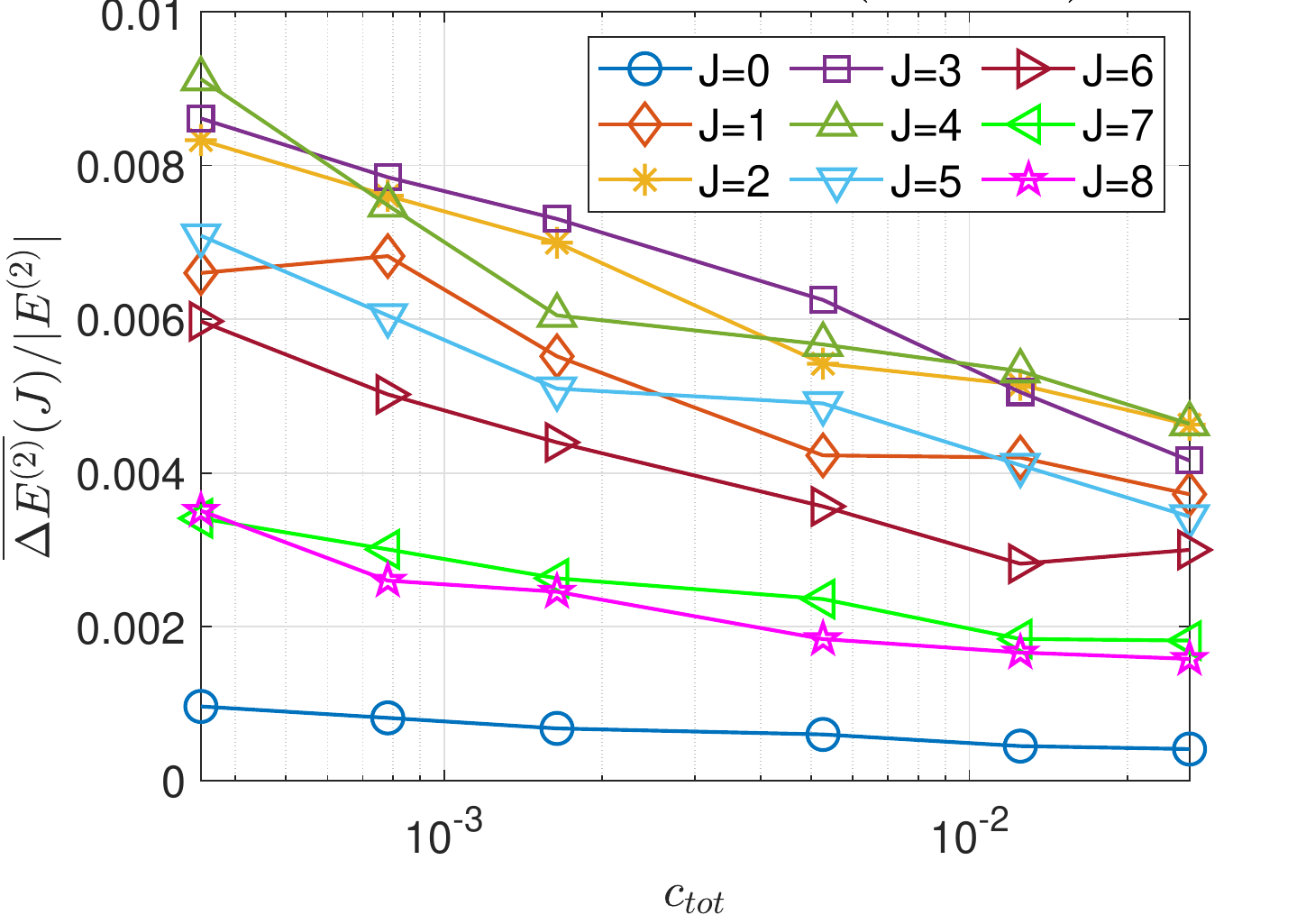}}
        \put(0.3250,0.0000){\includegraphics[width=0.41\unitlength]{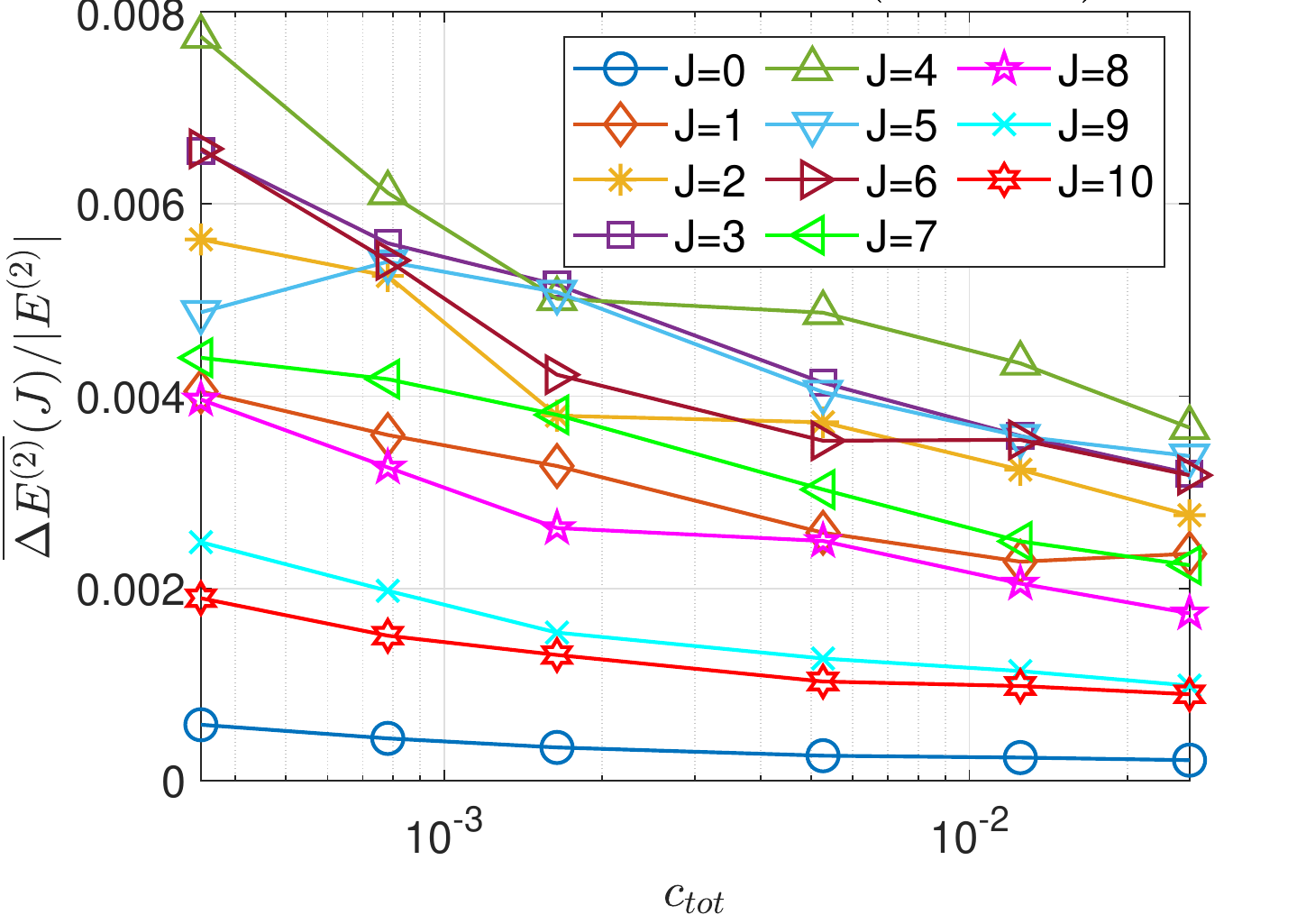}}
        \put(0.0200,0.8700){\fbox{\parbox{0.05\unitlength}{\centering \large $\nuc{Ca}{40}$}}}
        \put(0.5200,0.8700){\fbox{\parbox{0.05\unitlength}{\centering \large $\nuc{Ca}{48}$}}}
        \put(0.0200,0.5700){\fbox{\parbox{0.05\unitlength}{\centering \large $\nuc{Ni}{56}$}}}
        \put(0.5200,0.5700){\fbox{\parbox{0.05\unitlength}{\centering \large $\nuc{Ni}{78}$}}}
        \put(0.2700,0.2700){\fbox{\parbox{0.05\unitlength}{\centering \large $\nuc{Sn}{132}$}}}
    \end{picture}
    \vspace{-15pt}
    \caption{Contributions of each angular momentum channel to the mean relative error $\overline{\Delta E^{(2)}}/|E^{(2)}|$ as a function of the compression $\ctot$. Calculations were performed with the EM$1.8/2.0$ interaction in an $\eMax=14$ basis, and a two-stage JL embedding (RFD \text{+} RFD)\textsubscript{real} with $c_2=0.2$ has been used. 200 trials were performed.}
     \label{fig:deltaE_per-channel_vs-comp_appendix}
\end{figure*}

Figure \ref{fig:deltaE_per-channel_vs-comp_appendix} extends the results for $\overline{\Delta E^{(2)}}(J)/|E^{(2)|}$ of Fig.~\ref{fig:deltaE_per-channel_vs-comp} for nuclei beyond $\nuc{O}{16}$. The mean relative errors grow exponentially as we decrease $\ctot$ over a wide range, with some allowance for fluctuations due to the random sampling performed by the JLE (cf.~Sec.~\ref{sec:E2}).

\bibliographystyle{spphys}
\bibliography{refs}

\end{document}